\documentstyle{livrev}
\input epsf
\begin{document}


\title{Critical phenomena in gravitational collapse}

\author{Carsten Gundlach \\ Enrico Fermi Institute, University of
Chicago, \\ 5640 S Ellis Avenue, Chicago, IL 60637, USA \\ and \\
Faculty of Mathematical Studies, University of Southampton, \\
Highfield, Southampton SO17 1BJ, UK \footnote{current address}}

\date{16 Nov 1999}

\maketitle


\begin{abstract}

As first discovered by Choptuik, the black hole threshold in the space
of initial data for general relativity shows both surprising structure
and surprising simplicity. Universality, power-law scaling of the
black hole mass, and scale echoing have given rise to the term
``critical phenomena''. They are explained by the existence of exact
solutions which are attractors within the black hole threshold, that
is, attractors of codimension one in phase space, and which are
typically self-similar. This review gives an introduction to the
phenomena, tries to summarize the essential features of what is
happening, and then presents extensions and applications of this basic
scenario. Critical phenomena are of interest particularly for creating
surprising structure from simple equations, and for the light they
throw on cosmic censorship and the generic dynamics of general relativity.

\end{abstract}


\tableofcontents


\section{Introduction}


We briefly introduce the topic of this review article in two ways: by
definition, and in a historical context.


\subsection{Definition of the topic}


An isolated system in general relativity typically ends up in one of
three distinct kinds of final state. It either collapses to a black
hole, forms a stable star, or explodes and disperses, leaving empty
flat spacetime behind. The phase space of isolated gravitating systems
is therefore divided into basins of attraction. One cannot usually
tell into which basin of attraction a given data set belongs by any
other method than evolving it in time to see what its final state
is. The study of these invisible boundaries in phase space is the
subject of the relatively new field of critical collapse.

Ideas from dynamical systems theories provide a qualitative
understanding of the time evolution of initial data near any of these
boundaries. At the particular boundary between initial data that form
black holes and data that disperse, scale-invariance plays an
important role in the dynamics. This gives rise to a power law for the
black hole mass. Scale-invariance, universality and power-law behavior
suggest the name ``critical phenomena in gravitational collapse''. In
this review I cover critical collapse in the wider sense described
above, of beginning near a boundary in the space of initial data.

Critical phenomena in statistical mechanics and in gravitational
collapse share scale-invariant physics and the presence of a
renormalization group, but while the former involves statistical
ensembles, general relativity is deterministically described by PDEs.


\subsection{Historical introduction}


In 1987 Christodoulou, who was studying the spherically symmetric
Einstein-scalar model analytically
\cite{Christodoulou1,Christodoulou2,Christodoulou3,Christodoulou4,Christodoulou5},
suggested to Matt Choptuik, who was investigating the same system
numerically, the following question \cite{Choptuik94}: Consider a
generic smooth one-parameter family of asymptotically flat smooth
initial data, such that for large values of the parameter $p$ a black
hole is formed, and no black hole is formed for small $p$. If one
makes a bisection search for the critical value $p_*$ where a black
hole is just formed, does the black hole have finite or infinitesimal
mass?  After developing advanced numerical methods for this purpose,
Choptuik managed to give highly convincing numerical evidence that the
mass is infinitesimal. Moreover he found two totally unexpected
phenomena \cite{Choptuik92}: The first is the now famous scaling
relation
\begin{equation}
\label{power_law}
M \simeq C \, (p-p_*)^\gamma
\end{equation}
for the black hole mass $M$ in the limit $p\simeq p_*$ (but
$p>p_*$). Choptuik found $\gamma\simeq 0.37$. The second is the
appearance of a highly complicated, scale-periodic solution for $p
\simeq p_*$. The logarithmic scale period of this solution,
$\Delta\simeq 3.44$, is a second dimensionless number coming out of
the blue. As a third remarkable phenomenon, both the ``critical
exponent'' and ``critical solution'' are ``universal'', that is the
same for all one-parameter families ever investigated. Similar
phenomena to Choptuik's results were quickly found in other systems
too, suggesting that they were limited neither to scalar field matter
nor to spherical symmetry. Most of what is now understood in critical
phenomena is based on a mixture of analytical and numerical work.

Critical phenomena are arguably the most important contribution from
numerical relativity to new knowledge in general relativity to
date. At first researchers were intrigued by the appearance of a
complicated ``echoing'' structure and two mysterious dimensionless
numbers in the evolution of generic smooth initial data. Later it was
realized that critical collapse also provides a natural route to naked
singularities, and that it constitutes a new generic strong field
regime of classical general relativity, similar in universal
importance to the black hole end states of collapse.


\subsection{Plan of this review}


In order to give the reader a flavor of the original work on critical
phenomena, I describe Choptuik's results in some detail in Section
\ref{section:phenomena}. This is followed by a table of references to
the many other matter models (including vacuum gravity) in which
critical collapse has been investigated subsequently.

Complementary to this phenomenological approach, the next three
sections contain a systematic discussion. Section
\ref{section:basicscenario} describes the basic mechanism of critical
collapse.  Key concepts are borrowed from dynamical systems and
renormalization group theory. I introduce the relativistic notions of
scale-invariance and scale-periodicity, define the concept of a
critical solution, and sketch the calculation of the critical
exponent.  The following section \ref{section:extendedscenario}
contains both horizontal and vertical extensions to the basic picture
that are, in my mind, less central. The dividing line between this and
the previous section is therefore somewhat arbitrary. Section
\ref{section:otheraspects} groups together areas of current research
where results are still lacking or tentative.

The present paper is a revised and updated version of
\cite{Gundlach_critreview1}. The number of papers dedicated to
critical collapse since the work of Choptuik is now more than one
hundred, although not all are cited here. Previous review papers
include \cite{Horne_MOG,Bizon,Gundlach_Banach,BradyCai}. Choptuik's
own review article is \cite{Choptuik_review}. For an interesting
general review of the physics of scale-invariance, see \cite{Wilson}.


\section{The phenomena}
\label{section:phenomena}


In this section we present a phenomenological view of critical
collapse. We present in some detail the spherically symmetric scalar
field coupled to gravity, the model in which Choptuik first discovered
critical phenomena, and describe his findings. Then we give a brief
overview of the other systems that have been investigated since then.


\subsection{Case study: the spherically symmetric scalar field} 
\label{Choptuik_results}


The system in which Christodoulou and Choptuik have studied
gravitational collapse is the spherically symmetric massless,
minimally coupled scalar field. It has the advantage of simplicity,
while the scalar radiation propagating at the speed of light mimics
gravitational waves. We describe the system, and Choptuik's results.


\subsubsection{Spherical scalar field: definition of the system}


We consider a spherically symmetric, massless scalar field minimally
coupled to general relativity. The Einstein equations are
\begin{equation}
\label{scalar_stress_energy}
G_{ab} = 8 \pi \left(\nabla_a \phi \nabla_b \phi - {1\over 2} g_{ab}
\nabla_c \phi \nabla^c \phi\right)
\end{equation}
and the matter equation is
\begin{equation}
\nabla_a \nabla^a \phi = 0.
\end{equation}
Note that the matter equation of motion is contained within the
contracted Bianchi identities. Choptuik chose Schwarzschild-like
coordinates 
\begin{equation}
\label{tr_metric}
ds^2 = - \alpha^2(r,t) \, dt^2 + a^2(r,t) \, dr^2 + r^2 \, d\Omega^2,
\end{equation}
where $d\Omega^2 = d\theta^2 + \sin^2\theta \, d\varphi^2$ is the
metric on the unit 2-sphere. This choice of coordinates is defined by
the radius $r$ giving the surface area of 2-spheres as $4\pi r^2$, and by
$t$ being orthogonal to $r$ (polar-radial coordinates). One more
condition is required to fix the coordinate completely. Choptuik chose
$\alpha=1$ at $r=0$, so that $t$ is the proper time of the central
observer.

In the auxiliary variables
\begin{equation}
\Phi = \phi_{,r}, \qquad \Pi={a\over\alpha} \phi_{,t},
\end{equation}
the wave equation becomes a first-order system,
\begin{eqnarray}
\label{wave}
\Phi_{,t} & = &  \left({\alpha\over a}\Pi\right)_{,r}, \\
\Pi_{,t} & = & {1\over r^2} \left(r^2{\alpha\over a} \Phi\right)_{,r}.
\end{eqnarray}
In spherical symmetry there are four algebraically independent
components of the Einstein equations. Of these, one is a linear
combination of derivatives of the other and can be disregarded. The
other three contain only first derivatives of the metric, namely
$a_{,t}$, $a_{,r}$ and $\alpha_{,r}$. Choptuik chose to use the
equations giving $a_{,r}$ and $\alpha_{,r}$ for his numerical scheme,
so that only the scalar field is evolved, but the two metric
coefficients are calculated from the matter at each new time
step. (The main advantage of such a numerical scheme is its
stability.) These two equations are
\begin{eqnarray} 
\label{da_dr}
{a_{,r}\over a}  + {a^2 -1 \over 2r} - 2\pi r (\Pi^2 + \Phi^2) & = &
0, \\
\label{dalpha_dr}
{\alpha_{,r}\over \alpha}  - {a_{,r}\over a}  - {a^2 -1 \over r} &
= & 0,
\end{eqnarray}
and they are, respectively, the Hamiltonian constraint and the slicing
condition. These four first-order equations totally describe the
system. For completeness, we also give the remaining Einstein
equation,
\begin{equation}
\label{da_dt}
{a_{,t}\over \alpha}  =  4\pi r \Phi \Pi.
\end{equation}


\subsubsection{Spherical scalar field: the black hole threshold}


The free data for the system are the two functions $\Pi(r,0)$ and
$\Phi(r,0)$. (In spherical symmetry, there are no physical degrees of
freedom in the gravitational field.) Choptuik investigated
one-parameter families of such data by evolving the data for many
values each of the parameter, say $p$. He examined a number of
families in this way. Some simple examples of such families are
$\Phi(r,0)=0$ and a Gaussian for $\Pi(r,0)$, with the parameter $p$ taken
to be either the amplitude of the Gaussian, with the width and center
fixed, or the width, with position and amplitude fixed, or the
position, with width and amplitude fixed. For the amplitude
sufficiently small, with width and center fixed, the scalar field will
disperse, and for sufficiently large amplitude it will form a black
hole. Generic 1-parameter families behave in this way, but this is
difficult to prove in generality. Christodoulou showed for the
spherically symmetric scalar field system that data sufficiently weak
in a well-defined way evolve to a Minkowski-like spacetime
\cite{Christodoulou0a,Christodoulou3}, and that a class of
sufficiently strong data forms a black hole \cite{Christodoulou2}.

But what happens in between?  Choptuik found that in all 1-parameter
families of initial data he investigated he could make arbitrarily
small black holes by fine-tuning the parameter $p$ close to the black
hole threshold. An important fact is that there is nothing visibly
special to the black hole threshold. One cannot tell that one given
data set will form a black hole and another one infinitesimally close
will not, short of evolving both for a sufficiently long
time. ``Fine-tuning'' of $p$ to the black hole threshold proceeds by
bisection: Starting with two data sets one of which forms a black
hole, try a third one in between along some one-parameter family
linking the two, drop one of the old sets and repeat.

With $p$ closer to $p_*$, the spacetime varies on ever smaller
scales. The only limit was numerical resolution, and in order to push
that limitation further away, Choptuik developed numerical
techniques that recursively refine the numerical grid in spacetime
regions where details arise on scales too small to be resolved
properly. In the end, Choptuik could determine $p_*$ up to a relative
precision of $10^{-15}$, and make black holes as small as $10^{-6}$
times the ADM mass of the spacetime. The power-law scaling
(\ref{power_law}) was obeyed from those smallest masses up to black
hole masses of, for some families, $0.9$ of the ADM mass, that is,
over six orders of magnitude \cite{Choptuik94}. There were no families
of initial data which did not show the universal critical solution and
critical exponent. Choptuik therefore conjectured that $\gamma$ is the
same for all one-parameter families of smooth, asymptotically flat
initial data that depend smoothly on the parameter, and that the
approximate scaling law holds ever better for arbitrarily small
$p-p_*$.

Choptuik's results for individual 1-parameter families of data suggest
that there is a smooth hypersurface in the (infinite-dimensional)
phase space of smooth data which divides black hole from non-black
hole data. Let $P$ be any smooth scalar function on the space so that
$P=0$ is the black hole threshold. Then, for any choice of $P$, there
is a second smooth function $C$ on the space so that the black hole
mass as a function of the initial data is
\begin{equation}
M = \cases{CP^\gamma & for $P>0$ \cr 0 & for $P<0$}
\end{equation}
The entire unsmoothness at the black hole threshold is now captured by
the non-integer power. We should stress that this formulation of
Choptuik's mass scaling result is not even a conjecture, as we have
not stated on what function space it is supposed to
hold. Nevertheless, considering 1-parameter families of initial data
is only a tool for numerical investigations of the the
infinite-dimensional space of initial data, and a convenient way of
expressing analytic approximations.

Clearly a collapse spacetime which has ADM mass 1, but settles down to
a black hole of mass (for example) $10^{-6}$ has to show structure on
very different scales. The same is true for a spacetime which is as
close to the black hole threshold, but on the other side: the scalar
wave contracts until curvature values of order $10^{12}$ are reached
in a spacetime region of size $10^{-6}$ before it starts to
disperse. Choptuik found that all near-critical spacetimes, for all
families of initial data, look the same in an intermediate region,
that is they approximate one universal spacetime, which is also called
the critical solution. This spacetime is scale-periodic in the sense
that there is a value $t_*$ of $t$ such that when we shift the origin
of $t$ to $t_*$, we have
\begin{equation}
\label{tr_scaling}
Z(r,t) = Z\left(e^{n\Delta}r,e^{n\Delta}t\right)
\end{equation}
for all integer $n$ and for $\Delta\simeq 3.44$, and where $Z$ stands
for any one of $a$, $\alpha$ or $\phi$ (and therefore also for $r\Pi$
or $r\Phi$). The accumulation point $t_*$ depends on the family, but
the scale-periodic part of the near-critical solutions does not.

This result is sufficiently surprising to formulate it once more in a
slightly different manner. Let us replace $r$ and $t$ by a pair of
auxiliary variables such that one of them is the logarithm of an
overall spacetime scale. A simple example is
\begin{equation}
\label{x_tau}
x = -{r \over t-t_*}, \quad \tau = - \ln\left(-{t-t_*\over L}\right),
\quad t<t_*.
\end{equation}
($\tau$ has been defined so that it increases as $t$ increases and
approaches $t_*$ from below. It is useful to think of $r$,
$t$ and $L$ as having dimension length in units $c=G=1$, and of $x$
and $\tau$ as dimensionless.)  Choptuik's observation, expressed in
these coordinates, is that in any near-critical solution there is a
space-time region where the fields $a$, $\alpha$ and $\phi$ are well
approximated by their values in a universal solution, as
\begin{equation}
Z(x,\tau) \simeq Z_*(x,\tau),
\end{equation}
where the fields $a_*$, $\alpha_*$ and $\phi_*$ of the critical solution 
have the property
\begin{equation}
Z_*(x,\tau+\Delta) = Z_*(x,\tau).
\end{equation}
The dimensionful constants $t_*$ and $L$ depend on the particular
one-parameter family of solutions, but the dimensionless critical
fields $a_*$, $\alpha_*$ and $\phi_*$, and in particular their
dimensionless period $\Delta$, are universal.

The evolution of near-critical initial data starts resembling the
universal critical solution beginning at some length scale $Le^{-\tau}$
that is related (with some factor of order one) to the initial data scale.
A slightly supercritical and a slightly
subcritical solution from the same family (so that $L$ and $t_*$ are
the same) are practically indistinguishable until they have reached a
very small scale where the one forms an apparent horizon, while the
other starts dispersing. If a black hole is formed, its mass is
related (with a factor of order one) to this scale, and so we have for the
range $\Delta\tau$ of $\tau$ on which a near-critical solution
approximates the universal one
\begin{equation}
\label{duration}
\Delta\tau \simeq \gamma \ln|p-p_*| + {\rm const.},
\end{equation}
where the unknown factors of order one give rise to the unknown
constant.  As the critical solution is periodic in $\tau$ with period
$\Delta$ for the number $N$ of scaling ``echos'' that are seen we then
have the expression
\begin{equation}
N \simeq \Delta^{-1} \gamma \ln|p-p_*| + {\rm const.}
\end{equation}
Note that this holds for both supercritical and subcritical solutions.

Choptuik's results have been repeated by a number of other authors.
Gundlach, Price and Pullin \cite{GPP2} could verify the mass scaling
law with a relatively simple code, due to the fact that it holds even
quite far from criticality. Garfinkle \cite{Garfinkle} used the fact
that recursive grid refinement in near-critical solutions is not
required in arbitrary places, but that all refined grids are centered
on $(r=0,t=t_*)$, in order to use a simple fixed mesh refinement on a
single grid in double null coordinates: $u$ grid lines accumulate at
$u=0$, and $v$ lines at $v=0$, with $(v=0,u=0)$ chosen to coincide
with $(r=0,t=t_*)$. Hamad\'e and Stewart \cite{HamadeStewart} have
written an adaptive mesh refinement algorithm based on a double null
grid (but using coordinates $u$ and $r$), and report even higher resolution
than Choptuik. Their coordinate choice also allowed them to follow the
evolution beyond the formation of an apparent horizon.


\subsection{Other matter models}


Results similar to Choptuik's were subsequently found for a variety of
other matter models. In some of these, qualitatively new phenomena
were discovered, and we have reviewed this body of work by phenomena
rather than by matter models. The number of matter models is now so
large that a presentation by matter models is given only in the form
of Table \ref{table:mattermodels}. The second column specifies the
type of critical phenomena that is seen (compare Sections
\ref{subsection:typeI} and \ref{subsection:phasediagrams}). The next
column gives references to numerical evolutions of initial data, while
the last two columns give references to the semi-analytic approach.

Most models in the table are restricted to spherical symmetry, and
their matter content is described by a few functions of space (radius)
and time. Two models in the table are quite different, and therefore
particularly interesting. The axisymmetric vacuum model (see Section
\ref{subsection:axisymm}) is unique in going beyond spherical symmetry
nonperturbatively and in being vacuum rather than containing
matter. The fact that similar phenomena to Choptuik's were found in
that model strongly suggests that critical phenomena are not artifacts
of spherical symmetry or a specific matter model.

The second exceptional model, collisionless matter (Vlasov equation)
model is distinguished by having a much larger number of matter
degrees of freedom. Here, the matter content is described by a
function not only of space and time but also momentum. Remarkably, no
scaling phenomena of the kind seen in the scalar field were discovered
in numerical collapse simulations. Collisionless matter appears to
show a mass gap in critical collapse that depends on the initial
matter -- black hole formation turns on with a mass that is a large
part of the ADM mass of the initial data
\cite{ReinRendallSchaeffer}. Therefore universality is not observed
either. It is important to both confirm and further investigate this
phenomenology, in order to understand it better. The explanation may
be that the numerical precision was not high enough to find critical
phenomena, or they may be genuinely absent, perhaps because the space
of possible matter configurations is so much bigger than the space of
metrics in this case.

Critical collapse of a massless scalar field in spherical symmetry in
six spacetime dimensions was investigated in
\cite{GarfinkleCutlerDuncan}. Results are similar to four spacetime
dimensions.

Related results not listed in the table concern spherically symmetric
dust collapse. Here, the entire spacetime, the Tolman-Bondi solution,
is given in closed form from the initial velocity and density
profiles. Excluding shell crossing singularities, there is a ``phase
transition'' between initial data forming naked singularities at the
center and data forming black holes. Which of the two happens depends
only the leading terms in an expansion of the initial data around
$r=0$ \cite{Christodoulou0,Jhingan}. One could argue that this fact
also makes the matter model rather unphysical.


\begin{table}
\caption{An overview of numerical and semi-analytic work in critical collapse}
\label{table:mattermodels}
\begin{tabular}{l | l | l | l | l}
\hline
Matter model & Type of & Collapse  & Critical &
Perturbations \\ 
& phenomena & simulations & solution & \\
\hline
Perfect fluid &&& \\
-- $k=1/3$ & II &\cite{EvansColeman} &  CSS \cite{EvansColeman} &
\cite{KoikeHaraAdachi} \\
-- general $k$ & II & \cite{NeilsenChoptuik} &
 CSS \cite{Maison,NeilsenChoptuik} 
& \cite{Maison,KoikeHaraAdachi2}, \cite{Gundlach_critfluid2} \footnotemark[0] \\
\hline
Real scalar field &&&& \\
-- massless, min. coupled & II & \cite{Choptuik91,Choptuik92,Choptuik94} &
DSS \cite{Gundlach_Chop1} 
& \cite{Gundlach_Chop2}, \cite{MartinGundlach}\footnotemark[0] \\
-- massive & I & \cite{BradyChambersGoncalves} 
& oscillating \cite{SeidelSuen} & \\
& II & \cite{Choptuik94}
& DSS \cite{HaraKoikeAdachi,GundlachMartin}\footnotemark[1] 
& \cite{HaraKoikeAdachi,GundlachMartin}\footnotemark[1] \\ 
-- conformally coupled & II &
\cite{Choptuik92} & DSS  &   \\
\hline
2-d sigma model &&&& \\
-- complex scalar ($\kappa=0$) & II & \cite{Choptuik_pc} &
DSS \cite{Gundlach_Chop2}\footnotemark[2], \cite{HE1}\footnotemark[3]  &
\cite{Gundlach_Chop2}\footnotemark[2], \cite{HE2}\footnotemark[3]\\
-- axion-dilaton ($\kappa=1$) & II & 
\cite{HamadeHorneStewart} & 
CSS \cite{EardleyHirschmannHorne,HamadeHorneStewart} & 
\cite{HamadeHorneStewart} \\
-- scalar-Brans-Dicke ($\kappa>0$) & II & \cite{LieblingChoptuik,Liebling}
& CSS, DSS & \\
-- general $\kappa$ including $\kappa<0$ & II & & CSS, DSS \cite{HE3} & \cite{HE3} \\
\hline 
Massless scalar electrodynamics & II & \cite{HodPiran_charge} &
DSS \cite{GundlachMartin}\footnotemark[4] &
\cite{GundlachMartin}\footnotemark[4]  \\ 
\hline
$SU(2)$ Yang-Mills & I & \cite{ChoptuikChmajBizon} &
static \cite{BartnikMcKinnon} & \cite{LavrelashviliMaison} \\
& II & \cite{ChoptuikChmajBizon} & DSS \cite{Gundlach_EYM} 
& \cite{Gundlach_EYM} \\
& ``III'' & \cite{ChoptuikHirschmannMarsa} &
colored BH \cite{Bizon0,VolkovGaltsov} &
\cite{StraumannZhou,VolkovBrodbeckLavrelashviliStraumann} \\
$SU(2)$ Skyrme model & I & \cite{BizonChmaj} & static \cite{BizonChmaj}
& \cite{BizonChmaj} \\
& II & \cite{BizonChmajTabor}  
& static \cite{BizonChmajTabor}\footnotemark[5]  & \\
\hline
$SO(3)$ mexican hat & II & \cite{Liebling2} & DSS & \\
\hline
Axisymmetric vacuum & II & \cite{AbrahamsEvans,AbrahamsEvans2} & DSS &
\\
\hline 
Vlasov & none? & \cite{ReinRendallSchaeffer} && \\
\end{tabular}
\end{table}

\footnotetext[0]{Nonspherical perturbations.}

\footnotetext[1]{The critical solution and its perturbations for the
massive scalar field are asymptotic to those of the massless scalar.}

\footnotetext[2]{The (DSS) critical solution for the real massless scalar
field is also the critical solution for the complex scalar field. The
additional perturbations are all stable \cite{Gundlach_Chop2}.}

\footnotetext[3]{There is also a CSS solution \cite{HE1}, but it has
three unstable modes, not only one \cite{HE2}.}

\footnotetext[4]{The scalar electrodynamics critical solution is again
the real scalar field critical solution. Its perturbations are
those of the complex scalar field.}

\footnotetext[5]{The $SU(2)$ Skyrme and Yang-Mills models in spherical
symmetry have the same critical solution.}


\section{The basic scenario}
\label{section:basicscenario}


In this section we take a more abstract point of view and present the
general ideas underlying critical phenomena in gravitational collapse,
without reference to a specific system. This is useful, because these
ideas are really quite simple, and are best formulated in the language
of dynamical systems rather than general relativity.


\subsection{The dynamical systems picture}


We shall pretend that general relativity as an infinite-dimensional
dynamical system. The phase space is the space of pairs of
three-metrics and extrinsic curvatures (plus any matter variables)
that obey the Hamiltonian and momentum constraints. In the following
we restrict ourselves to asymptotically flat data. In other words, it
is the space of initial data for an isolated self-gravitating
system. The evolution equations are the ADM equations. They contain
the lapse and shift as free fields that can be given arbitrary
values. In order to obtain an autonomous dynamical system, one needs a
general prescription that provides a lapse and shift for given initial
data. What such a prescription could be is very much an open problem
and is discussed below in section \ref{subsection:coordinates}. That
is the first gap in the dynamical systems picture. The second gap is
that even with a prescription for the lapse and shift in place, a
given spacetime does not correspond to a unique trajectory in phase
space, but to many, depending on how the spacetime is sliced. A
possibility would be to restrict the phase space further, for example
to maximal slices only. The third problem is that in order to talk
about attractors and repellers we need a notion of convergence on the
phase space, that is a distance measure. In the following, we brazenly
ignore all three gaps in order to apply some fundamental concepts of
dynamical systems theory to gravitational collapse.

An isolated system in general relativity, such as a star, or ball of
radiation fields, or even of pure gravitational waves, typically ends
up in one of three kinds of final state. It either collapses to a
black hole, forms a stable star, or explodes and disperses, leaving
empty flat spacetime behind. The phase space of isolated gravitating
systems is therefore divided into basins of attraction. A boundary
between two basins of attraction is called a critical surface.  All
numerical results are consistent with the idea that these boundaries
are smooth hypersurfaces of codimension one in the phase space of GR.
Inside the dispersion basin, Minkowski spacetime is an attractive
fixed point.  Inside the black hole basin, the 3-parameter family of
Kerr-Newman black holes forms a manifold of attracting fixed points.
(Clearly these are attractors only in a distance measures that uses
amplitude rather than total energy for waves traveling off to
infinity.)

A phase space trajectory starting in a critical surface by definition
never leaves it. A critical surface is therefore a dynamical system in
its own right, with one dimension fewer. Say it has an attracting fixed
point or attracting limit cycle. For the black hole threshold in all
toy models that have been examined (with the possible exception of the
Vlasov-Einstein system, see Section \ref{section:conclusions} below)
this is the case. We shall call these a critical point, or critical
solution, or critical spacetime. Within the complete phase space, the
critical solution is an attractor of codimension one. It has an
infinite number of decaying perturbation modes tangential to the
critical surface, and a single growing mode that is not tangential.

\begin{figure}
\label{fig:dynsim}
\epsfysize=10cm
\centerline{\epsffile{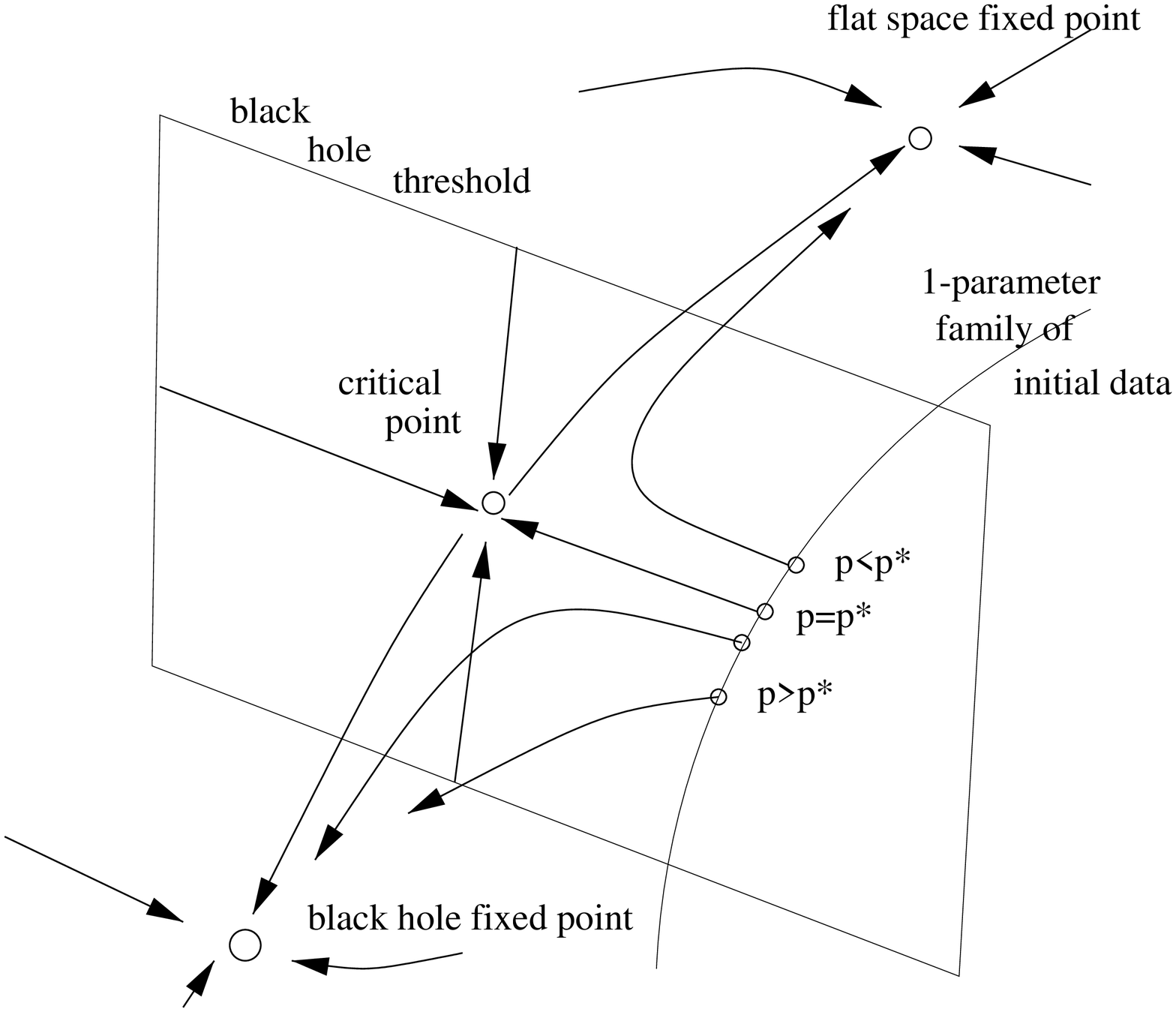}}
\caption{The phase space picture for the black hole-dispersion
threshold in the presence of a CSS solution. The arrow lines are time
evolutions, corresponding to spacetimes. The line without an arrow is
not a time evolution, but a 1-parameter family of initial data that
crosses the black hole threshold at $p=p_*$.}
\end{figure}

\begin{figure}
\label{fig:phasespace}
\epsfysize=10cm
\centerline{\epsffile{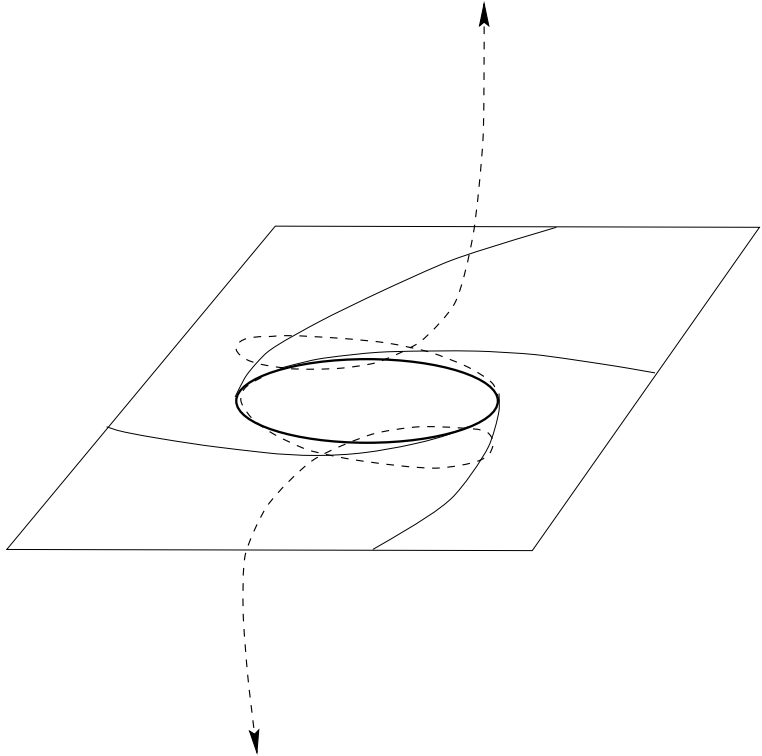}}
\caption{The phase space picture for discrete
self-similarity. The plane represents the critical surface. (In
reality this is a hypersurface of co-dimension one in an
infinite-dimensional space.) The circle (fat unbroken line) is the
limit cycle representing the critical solution. The thin unbroken
curves are spacetimes attracted to it. The dashed curves are
spacetimes repelled from it. There are two families of such curves,
labeled by one periodic parameter, one forming a black hole, the other
dispersing to infinity. Only one member of each family is shown.}
\end{figure}

Any trajectory beginning near the critical surface, but not
necessarily near the critical point, moves almost parallel to the
critical surface towards the critical point. As the critical point is
approached, the parallel movement slows down, and the phase point
spends some time near the critical point. Then the phase space point
moves away from the critical point in the direction of the growing
mode, and ends up on a fixed point. This is the origin of
universality: any initial data set that is close to the black hole
threshold (on either side) evolves to a spacetime that approximates
the critical spacetime for some time. When it finally approaches
either empty space or a black hole it does so on a trajectory appears
to be coming from the critical point itself. All near-critical
solutions are passing through one of these two funnels. All details of
the initial data have been forgotten, except for the distance from the
black hole threshold.  The phase space picture in the presence of a
fixed point critical solution is sketched in
Fig. \ref{fig:dynsim}. The phase space picture in the presence of a
limit cycle critical solution is sketched in
Fig. \ref{fig:phasespace}.

All critical points that have been found in black hole thresholds so
far have an additional symmetry, either continuous or discrete. They
are either time-independent (static) or periodic in time, or
scale-independent of scale-periodic (discretely or continuously
self-similar). The static or periodic critical points are metastable
stars. As we shall see below in section \ref{subsection:typeI}, they
give rise to a finite mass gap at the black hole threshold. In the
remainder of this section we concentrate on the self-similar fixed
points. They give rise to power-law scaling of the black hole mass at
the threshold. These are the phenomena discovered by Choptuik. They
are now referred to as type II critical phenomena, while the type with
the mass gap, historically discovered second, is referred to as type
I.

Continuously scale-invariant, or self-similar, solutions arise as
intermediate attractors in some fluid dynamics problems (without
gravity) \cite{BarenblattZeldovich,Barenblatt,Barenblatt2}. Discrete
self-similarity does not seem to have played a role in physics before
Choptuik's discoveries.

It is clear from the dynamical systems picture that the closer the
initial phase point (data set) is to the critical surface, the closer
the phase point will get to the critical point, and the longer it will
remain close to it. Making this observation quantitative will give rise
to Choptuik's mass scaling law in section \ref{subsection:scaling}
below. But we first need to define self-similarity in GR.


\subsection{Scale-invariance and self-similarity}


The critical solution found by Choptuik
\cite{Choptuik91,Choptuik92,Choptuik94} for the spherically symmetric
scalar field is scale-periodic, or discretely self-similar (DSS),
while other critical solutions, for example for a spherical perfect
fluid \cite{EvansColeman} are scale-invariant, or continuously
self-similar (CSS). We begin with the continuous symmetry because it
is simpler. In Newtonian physics, a solution $Z$ is self-similar if it
is of the form
\begin{equation}
Z(\vec x, t) = Z\left[{\vec x\over f(t)}\right]
\end{equation}
If the function $f(t)$ is derived from dimensional considerations
alone, one speaks of self-similarity of the first kind. An example is
$f(t)=\sqrt{\lambda t}$ for the diffusion equation $Z_{,t}=\lambda
\Delta Z$. In more complicated equations, the limit of self-similar
solutions can be singular, and $f(t)$ may contain additional
dimensionful constants (which do not appear in the field equation) in
terms such as $(t/L)^\alpha$, where $\alpha$, called an anomalous
dimension, is not determined by dimensional considerations but through
the solution of an eigenvalue problem \cite{Barenblatt}.  

A continuous self-similarity of the spacetime in GR corresponds to the
existence of a homothetic vector field $\xi$, defined by the property
\cite{CahillTaub}
\begin{equation}
\label{homothetic_metric}
{\cal L}_\xi g_{ab} = 2 g_{ab}.
\end{equation}
This is a special type of conformal Killing vector, namely one with
constant coefficient on the right-hand side. The value of
this constant coefficient is conventional, and can be set equal to 2 by
a constant rescaling of $\xi$. From (\ref{homothetic_metric}) it
follows that
\begin{equation}
\label{homothetic_curvature}
{\cal L}_\xi {R^a}_{bcd} = 0,
\end{equation}
and therefore
\begin{equation}
\label{homothetic_matter}
{\cal L}_\xi G_{ab} = 0,
\end{equation}
but the inverse does not hold: the Riemann tensor and the metric need
not satisfy (\ref{homothetic_curvature}) and (\ref{homothetic_metric})
if the Einstein tensor
obeys (\ref{homothetic_matter}). If the matter is a perfect fluid
(\ref{fluid_stress_energy}) it follows from (\ref{homothetic_metric}),
(\ref{homothetic_matter}) and the Einstein equations that
\begin{equation}
{\cal L}_\xi u^a = -u^a, \quad {\cal L}_\xi \rho = -2\rho, \quad
{\cal L}_\xi p = -2p.
\end{equation}
Similarly, if the matter is a massless scalar field $\phi$, with
stress-energy tensor (\ref{scalar_stress_energy}), it follows that
\begin{equation}
\label{scalar_CSS}
{\cal L}_\xi \phi=\kappa, 
\end{equation}
where $\kappa$ is a constant. 

In coordinates $x^\mu=(\tau,x^i)$ adapted to the homothety, the metric
coefficients are of the form
\begin{equation}
\label{CSS_coordinates}
g_{\mu\nu}(\tau,x^i) = e^{-2\tau} \tilde g_{\mu\nu}(x^i)
\end{equation}
where the coordinate $\tau$ is the negative logarithm of a
spacetime scale, and the remaining three coordinates $x^i$ are
dimensionless. In these coordinates, the homothetic vector field is
\begin{equation}
\label{xi_in_coordinates}
\xi = - {\partial\over\partial\tau}.
\end{equation}
The minus sign in both equations (\ref{CSS_coordinates}) and
(\ref{xi_in_coordinates}) is a convention we have chosen so that
$\tau$ increases towards smaller spacetime scales. For the critical
solutions of gravitational collapse, we shall later choose surfaces of
constant $\tau$ to be spacelike (although this is not possible
globally), so that $\tau$ is the time coordinate as well as the scale
coordinate. Then it is natural that $\tau$ increases towards the
future, that is towards smaller scales.

As an illustration, the CSS scalar field in these coordinates would be
\begin{equation}
\phi=f(x)+\kappa\tau,
\end{equation}
with $\kappa$ a constant. Similarly, perfect fluid matter with
stress-energy 
\begin{equation}
\label{fluid_stress_energy}
G_{ab} = 8 \pi \left[(p+\rho) u_a u_b + p g_{ab}\right]
\end{equation}
with the scale-invariant equation of state $p=k\rho$, $k$ a constant,
allows for CSS solutions where
the direction of $u^a$ depends only on $x$, and the density is of the
form
\begin{equation}
\rho(x,\tau) = e^{2\tau} f(x).
\end{equation}

The generalization to a discrete self-similarity is
obvious in these coordinates, and was made in \cite{Gundlach_Chop2}:
\begin{equation}
\label{DSS_coordinates}
g_{\mu\nu}(\tau,x^i) = e^{-2\tau} \tilde g_{\mu\nu}(\tau,x^i), \quad
\hbox{where} \quad \tilde g_{\mu\nu}(\tau,x^i) = \tilde 
g_{\mu\nu}(\tau+\Delta,x^i).
\end{equation}
The conformal metric $\tilde g_{\mu\nu}$ does now depend on $\tau$,
but only in a periodic manner. Like the continuous symmetry, the
discrete version has a geometric formulation
\cite{Gibbonspc}: A spacetime is discretely self-similar if 
there exists a discrete diffeomorphism $\Phi$ and a real constant
$\Delta$ such that
\begin{equation} 
\label{DSS_geometric}
\Phi^* g_{ab} = e^{2\Delta} g_{ab},
\end{equation}
where $\Phi^* g_{ab}$ is the pull-back of $g_{ab}$ under the
diffeomorphism $\Phi$. This is our definition of discrete
self-similarity (DSS). It can be obtained formally from
(\ref{homothetic_metric}) by integration along $\xi$ over an interval
$\Delta$ of the affine parameter. Nevertheless, the definition is
independent of any particular vector field $\xi$. One simple
coordinate transformation that brings the Schwarzschild-like
coordinates (\ref{tr_metric}) into the form (\ref{DSS_coordinates})
was given in Eqn. (\ref{x_tau}), as one easily verifies by
substitution.  The most general ansatz for the massless scalar field
compatible with DSS is
\begin{equation}
\label{scalar_DSS}
\phi=f(\tau,x^i) + \kappa \tau, \quad {\rm where} \quad
f(\tau,x^i)=f(\tau+\Delta,x^i)
\end{equation}
with $\kappa$ a constant. (In the Choptuik critical solution,
$\kappa=0$ for unknown reasons.)

It should be stressed here that the coordinate systems adapted to CSS
(\ref{CSS_coordinates}) or DSS (\ref{DSS_coordinates}) form large
classes, even in spherical symmetry. One can fix the surface $\tau=0$
freely, and can introduce any coordinates $x^i$ on it. In particular,
in spherical symmetry, $\tau$-surfaces can be chosen to be spacelike,
as for example defined by (\ref{tr_metric}) and (\ref{x_tau}) above,
and in this case the coordinate system cannot be global (in the
example, $t<0$).  Alternatively, one can find global coordinate
systems, where $\tau$-surfaces must become spacelike at large $r$, as
in the coordinates (\ref{global_CSS}). Moreover, any such coordinate
coordinate system can be continuously deformed into one of the same
class.

In a possible source of confusion, Evans and Coleman
\cite{EvansColeman} use the term ``self-similarity of the second
kind'', because they define their self-similar coordinate $x$ as
$x=r/f(t)$, with $f(t)=t^n$. Nevertheless, the spacetime they
calculate is homothetic, or ``self-similar of the first kind''
according to the terminology of Carter and Henriksen
\cite{CarterHenriksen,Coley} . The difference is only a coordinate
transformation: the $t$ of \cite{EvansColeman} is not proper time at
the origin, but what would be proper time at infinity if the spacetime
was truncated at finite radius and matched to an asymptotically flat
exterior \cite{Evanspc}.

There is a large body of research on spherically symmetric
self-similar perfect fluid solutions
\cite{CarrColey,Bogoyavlenskii,FoglizzoHenriksen,BicknellHenriksen,OriPiran,OriPiran2,LakeZannias}. Scalar
field spherically symmetric CSS solutions were examined in
\cite{GoldwirthPiran,Brady_CSS_scalar}. In these papers, the Einstein
equations are reduced to an ODE system by the self-similar spherically
symmetric ansatz, which is then discussed as a dynamical
system. Surprisingly, the critical solutions of gravitational collapse
were explicitly constructed only once they had been seen in collapse
simulations. The critical solution found in perfect fluid collapse
simulations was constructed through a CSS ansatz by Evans and Coleman
\cite{EvansColeman}. In this ansatz, the requirement of analyticity at
the center and at the past matter characteristic of the singularity
provides sufficient boundary conditions for the ODE system. (For
claims to the contrary see
\cite{CarrColeyGoliathNilssonUggla,CarrHenriksen}.  The DSS scalar
critical solution of scalar field collapse was constructed by Gundlach
\cite{Gundlach_Chop1,Gundlach_Chop2} using a similar method. More
details of how the critical solutions are constructed using a DSS or
CSS ansatz are discussed in Section \ref{subsection:regular}.


\subsection{Black hole mass scaling} 
\label{subsection:scaling}


The following calculation of the critical exponent from the linear
perturbations of the critical solution by dimensional analysis was
suggested by Evans and Coleman \cite{EvansColeman} and carried out by
Koike, Hara and Adachi \cite{KoikeHaraAdachi} and Maison
\cite{Maison}. It was generalized to the discretely self-similar (DSS)
case by Gundlach \cite{Gundlach_Chop2}. For simplicity of notation we
consider again the spherically symmetric CSS case.  The DSS case is
discussed in \cite{Gundlach_Chop2}.

Let $Z$ stand for a set of scale-invariant variables of the problem in
a first-order formulation. $Z(r)$ is an element of the phase space,
and $Z(r,t)$ a solution. The self-similar solution is of the form
$Z(r,t)=Z_*(x)$.  In the echoing region, where $Z_*$ dominates, we
linearize around it. As the background solution is $\tau$-independent,
$Z(x,\tau)=Z_*(x)$, its linear perturbations can depend on $\tau$ only
exponentially (with complex exponent $\lambda$), that is
\begin{equation}
\delta Z(x,\tau)=\sum_{i=1}^\infty C_i \, e^{\lambda_i \tau} Z_i(x),
\end{equation}
where the $C_i$ are free constants.  To linear order, the solution
in the echoing region is then of the form
\begin{equation}
Z(x,\tau;p) \simeq Z_*(x) + \sum_{i=1}^\infty C_i(p) \, e^{\lambda_i
\tau} Z_i(x).
\end{equation} 
The coefficients $C_i$ depend in a complicated way on the initial
data, and hence on $p$. If $Z_*$ is a critical solution, by definition
there is exactly one $\lambda_i$ with positive real part (in fact it
is purely real), say $\lambda_0$. As $t\to t_*$ from below and
$\tau\to\infty$, all other perturbations vanish. In the following we
consider this limit, and retain only the one growing perturbation. By
definition the critical solution corresponds to $p=p_*$, so we must
have $C_0(p_*)=0$. Linearizing around $p_*$, we obtain
\begin{equation}
\label{echoing_region}
\lim_{\tau\to\infty} Z(x,\tau) \simeq Z_*(x) + {dC_0\over dp} (p-p_*)
e^{\lambda_0\tau} Z_0(x).
\end{equation}
This approximate solution explains why the solution $Z_*$ is
universal.  It is now also clear why Eqn. (\ref{duration}) holds, that
is why we see more of the universal solutions (in the DSS case, more
``echos'') as $p$ is tuned closer to $p_*$. The critical solution
would be revealed up to the singularity $\tau=\infty$ if perfect
fine-tuning of $p$ was possible. A possible source of confusion is
that the critical solution, because it is self-similar, is not
asymptotically flat. Nevertheless, it can arise in a region up to
finite radius as the limiting case of a family of asymptotically flat
solutions. At large radius, it is matched to an asymptotically flat
solution which is not universal but depends on the initial data (as
does the place of matching.)

The solution has the approximate form (\ref{echoing_region}) over a
range of $\tau$. Now we extract Cauchy data at one particular value of
$\tau$ within that range, namely $\tau_*$ defined by
\begin{equation}
{dC_0\over dp} (p-p_*) e^{-\lambda_0\tau_*} \equiv \epsilon,
\end{equation}
where $\epsilon$ is some constant $\ll 1$, so that at this $\tau$ the
linear approximation is still valid. Note that $\tau_*$ depends on
$p$. At sufficiently large $\tau$, the linear perturbation has grown
so much that the linear approximation breaks down. Later on a black
hole forms.  The crucial point is that we need not follow this
evolution in detail, nor does it matter at what amplitude $\epsilon$
we consider the perturbation as becoming non-linear. It is sufficient
to note that the Cauchy data at $\tau=\tau_*$ depend on $r$ only
through the argument $x$, because by definition of $\tau_*$ we have
\begin{equation}
Z(x,\tau_*) \simeq Z_*(x) + \epsilon \ Z_0(x).
\end{equation}
Going back to coordinates $t$ and $r$ we have
\begin{equation}
Z(r,t_*-L_*) \simeq Z_*\left(-{r\over L_*}\right) + \epsilon \
Z_0\left(-{r\over L_*} \right), \qquad L_* \equiv Le^{-\tau_*}.
\end{equation}
These intermediate data at $t=t_*$ depend on the initial data at $t=0$
only through the overall scale $L_*$.  The field equations themselves
do not have an intrinsic scale. It follows that the solution based on
the data at $t_*$ must be universal up to the overall scale. In
suitable coordinates (for example the polar-radial coordinates of
Choptuik) it is then of the form
\begin{equation}
Z(r,t) = f\left({r\over L_*}, {t-t_*\over L_*}\right),
\end{equation}
for some function $f$ that is universal for all 1-parameter families
\cite{HE2}. This universal form of the solution applies for all
$t>t_*$, even after the approximation of linear perturbation theory
around the critical solution breaks down. Because the black hole mass
has dimension length, it must be proportional to $L_*$, the only
length scale in the solution. Therefore
\begin{equation}
M \propto L_* \propto (p-p_*)^{1\over \lambda_0},
\end{equation}
and we have found the critical exponent $\gamma = 1/\lambda_0$.

When the critical solution is DSS, the scaling law is modified.  This
was predicted in \cite{Gundlach_Chop2} and predicted independently and
verified in collapse simulations by Hod and Piran
\cite{HodPiran_wiggle}. On the straight line relating $\ln M$ to
$\ln(p-p_*)$, a periodic ``wiggle'' or ``fine structure'' of small
amplitude is superimposed:
\begin{equation}
\ln M = \gamma \ln (p-p_*) + c + f[\gamma \ln (p-p_*) + c],
\end{equation}
with $f(z)=f(z+\Delta)$.  The periodic function $f$ is again universal
with respect to families of initial data, and there is only one
parameter $c$ that depends on the family of initial data,
corresponding to a shift of the wiggly line in the $\ln(p-p_*)$
direction. (No separate adjustment in the $\ln M$ direction is
possible.)

It is easy to see that the maximal value of the scalar curvature, and
similar quantities, for near-critical solutions, scale just like the
black hole mass, with a critical exponent $2\gamma$. Technically, it
is easier to measure the critical exponent and the fine-structure in
the subcritical regime from the maximum curvature than from the black
hole mass in the supercritical regime \cite{GarfinkleDuncan}.


\section{Extensions of the basic scenario}
\label{section:extendedscenario}


In the previous section we have tried to present the {\it central}
ideas of critical collapse. Much more is now known however. In this
section we present other aspects that are either horizontal or
vertical extensions of the central ideas.


\subsection{Black hole thresholds with a mass gap} \label{subsection:typeI}


The spherical $SU(2)$ Einstein-Yang-Mills system
\cite{ChoptuikChmajBizon,BizonChmaj,BizonChmaj2,BizonChmaj3} shows two
different kinds of critical phenomena, dominated by two different
critical solutions. Which kind of behavior arises appears to depend on
the qualitative shape of the initial data. In one kind of behavior,
black hole formation turns on at an infinitesimal mass with the
familiar power-law scaling, dominated by a DSS critical solution. In
the other kind, black hole formation turns on at a finite mass, and
the critical solution is now a static, asymptotically flat solution
which had been found before by Bartnik and McKinnon
\cite{BartnikMcKinnon}. Choptuik, Chmaj and Bizon labelled the two
kinds of critical behavior type II and type I respectively,
corresponding to a second- and a first-order phase transition. The
newly found, type I critical phenomena show a scaling law that is
mathematically similar to the black hole mass scaling observed in type
II critical phenomena.  Let $\partial/\partial t$ be the static
Killing vector of the critical solution. Then the perturbed critical
solution is of the form
\begin{equation}
\label{typeIintermediate}
Z(r,t) = Z_*(r) + {d C_0\over dp} (p-p_*) e^{\lambda_0 t} Z_0(r) +
\hbox{decaying modes}.
\end{equation}
This is similar to Eqn. (\ref{echoing_region}), but the growth of the
unstable mode is now exponential in $t$, not in $\ln t$. In a close
parallel to $\tau_*$, we define a time $t_p$ by
\begin{equation}
{d C_0\over dp} (p-p_*) e^{\lambda_0 t_p} \equiv \epsilon,
\end{equation}
so that the initial data at $t_p$ are
\begin{equation}
Z(r,t_p) \simeq Z_*\left(r\right) + \epsilon \ Z_0\left(r \right),
\end{equation}
and so the final black hole mass is independent of $p-p_*$. (It is of
the order of the mass of the static critical solution.)  The scaling
is only apparent in the lifetime of the critical solution, which we
can take to be $t_p$. It is
\begin{equation}
t_p = - {1\over \lambda_0} \ln(p-p_*) + {\rm const.}
\end{equation}

The type I critical solution can also have a discrete symmetry, that
is, can be periodic in time instead of being static. This behavior was
found in collapse situations of the massive scalar field by Brady,
Chambers and Gon\c calves \cite{BradyChambersGoncalves}. Previously,
Seidel and Suen \cite{SeidelSuen} had constructed periodic,
asymptotically flat, spherically symmetric self-gravitating massive
scalar field solutions they called oscillating soliton stars. By
dimensional analysis, the scalar field mass $m$ sets an overall scale
of $1/m$ (in units $G=c=1$). For given $m$, Seidel and Suen found a
one-parameter family of such solutions with two branches. The more
compact solution for a given ADM mass is unstable, while the more
extended one is stable to spherical perturbations. Brady, Chambers and
Gon\c calves (BCG) report that the type I critical solutions they find
are from the unstable branch of the Seidel and Suen solutions.  They
see a one-parameter family of (type I) critical solutions, rather than
an isolated critical solution. BCG in fact report that the black hole
mass gap does depend on the initial data.  As expected from the
discrete symmetry, they find a small wiggle in the mass of the
critical solution which is periodic in $\ln(p-p_*)$. If type I or type
II behavior is seen appears to depend mainly on the ratio of the
length scale of the initial data to the length scale $1/m$.

In the critical phenomena that were first observed, with an isolated
critical solution, only one number's worth of information, namely the
separation $p-p_*$ of the initial data from the black hole threshold,
survives to the late stages of the time evolution. Recall that our
definition of a critical solution is one that has exactly one unstable
perturbation mode, with a black hole formed for one sign of the
unstable mode, but not for the other. This definition does not exclude
an $n$-dimensional family of critical solutions. Each solution in the
family then has $n$ marginal modes leading to neighboring critical
solutions, as well as the one unstable mode. $n+1$ numbers' worth of
information survive from the initial data, and the mass gap in type I,
or the critical exponent for the black hole mass in type II, for
example, depend on the initial data through $n$ parameters. In other
words, universality exists in diminished form. The results of BCG are
an example of a one-parameter family of type I critical
solutions. Recently, Brodbeck et al. \cite{Brodbecketal} have shown,
under the assumption of linearization stability, that there is a
one-parameter family of stationary, rotating solutions beginning at
the (spherically symmetric) Bartnik-McKinnon solution. This could turn
out to be a second one-parameter family of type I critical solutions,
provided that the Bartnik-McKinnon solution does not have any unstable
modes outside spherical symmetry (which has not yet been investigated)
\cite{Rendallpc}.

Bizo\'n and Chmaj have studied type I critical collapse of an $SU(2)$
Skyrme model coupled to gravity, which in spherical symmetry with a
hedgehog ansatz is characterized by one field $F(r,t)$ and one
dimensionless coupling constant $\alpha$. Initial data $F(r)\sim
\tanh(r/p)$, $\dot F(r)=0$ surprisingly form black holes for both
large and small values of the parameter $p$, while for an intermediate
range of $p$ the endpoint is a stable static solution called a
skyrmion. (If $F$ was a scalar field, one would expect only one
critical point on this family.) The ultimate reason for this behavior
is the presence of a conserved integer ``baryon number'' in the matter
model. Both phase transitions along this one-parameter family are
dominated by a type I critical solution, that is a different skyrmion
which has one unstable mode. In particular, an intermediate time
regime of critical collapse evolutions agrees well with an ansatz of
the form (\ref{typeIintermediate}), where $Z_*$, $Z_0$ and $\lambda$
were obtained independently. It is interesting to note that the type I
critical solution is singular in the limit $\alpha \to 0$, which is
equivalent to $G \to 0$, because the known type II critical solutions
for any matter model also do not have a weak gravity limit.

Apparently, type I critical phenomena can arise even without the
presence of a scale in the field equations. A family of exact
spherically symmetric, static, asymptotically flat solutions of vacuum
Brans-Dicke gravity given by van Putten was found by Choptuik,
Hirschmann and Liebling \cite{ChopHirschLieb} to sit at the black
hole-threshold and to have exactly one growing mode. This family has
two parameters, one of which is an arbitrary overall scale.


\subsection{CSS and DSS critical solutions}


Critical solutions are continuously or discretely self-similar, and
have exactly one growing perturbation mode. Other regular CSS or DSS
solutions have more than one growing mode, and so will not appear as
critical solution at the black hole threshold. An example for this is
provided by the spherically symmetric massless complex scalar field.
Hirschmann and Eardley \cite{HE1} found a way of constructing a CSS
scalar field solution by making the scalar field $\phi$ complex but
limiting it to the ansatz
\begin{equation}
\phi=e^{i\omega\tau} f(x),
\end{equation}
with $\omega$ a real constant and $f$ real. The metric is then
homothetic, while the scalar field shows a trivial kind of ``echoing''
in the complex phase. Later, they found that this solution has three
modes with ${\rm Re}\lambda>0$ \cite{HE2} and is therefore not the
critical solution. On the other hand, Gundlach \cite{Gundlach_Chop2}
examined complex scalar field perturbations around Choptuik's real
scalar field critical solution and found that only one of them, purely
real, has ${\rm Re}\lambda>0$, so that the real scalar field critical
solution is a critical solution (up to an overall complex phase) also
for the free complex scalar field. This had been seen already in
collapse calculations \cite{Choptuik_pc}.

As the symmetry of the critical solution, CSS or DSS, depends on the
matter model, it is interesting to investigate critical behavior in
parameterized families of matter models. Two such one-parameter
families have been investigated. The first one is the spherical
perfect fluid with equation of state $p=k\rho$ for arbitrary
$k$. Maison \cite{Maison} constructed the regular CSS solutions and
its linear perturbations for a large number of values of $k$. In each
case, he found exactly one growing mode, and was therefore able to
predict the critical exponent. (To my knowledge, these critical
exponents have not yet been verified in collapse simulations.) As Ori
and Piran before \cite{OriPiran,OriPiran2}, he claimed that there are
no regular CSS solutions for $k> 0.88$. Recently, Neilsen and Choptuik
\cite{NeilsenChoptuik,NeilsenChoptuik2} have found CSS critical
solutions for all values of $k$ right up to $1$, both in collapse
simulations and by making a CSS ansatz. Interesting questions arise
because the stiff ($p=\rho$) perfect fluid, limited to irrotational
solutions, is equivalent to the massless scalar field, limited to
solutions with timelike gradient, while the scalar field critical
solution is actually DSS. These are currently being investigated
\cite{BradyGundlachNeilsen}.

The second one-parameter family of matter models was suggested by
Hirschmann and Eardley \cite{HE3}, who looked for a natural way of
introducing a non-linear self-interaction for the (complex) scalar
field without introducing a scale. (We discuss dimensionful coupling
constants in the following sections.) They investigated the model
described by the action
\begin{equation}
S=\int \sqrt{g}\left(R-{2|\nabla\phi|^2\over
(1-\kappa|\phi|^2)^2}\right).
\end{equation}
Note that $\phi$ is now complex, and the parameter $\kappa$ is real
and dimensionless. This is a 2-dimensional sigma model with a target
space metric of constant curvature (namely $\kappa$), minimally
coupled to gravity. Moreover, for $\kappa > 0$ there are (nontrivial)
field redefinitions which make this model equivalent to a real
massless scalar field minimally coupled to Brans-Dicke gravity, with
the Brans-Dicke coupling given by
\begin{equation}
\omega_{\rm BD}=-{3\over2}+{1\over 8\kappa}.
\end{equation}
In particular, $\kappa=1$ ($\omega_{\rm BD}=-11/8$) corresponds to an
axion-dilaton system arising in string theory
\cite{EardleyHirschmannHorne}. $\kappa=0$ is the free complex scalar
field coupled to Einstein gravity).  Hirschmann and Eardley calculated
a CSS solution and its perturbations and concluded that it is the
critical solution for $\kappa>0.0754$, but has three unstable modes
for $\kappa<0.0754$. For $\kappa<-0.28$, it acquires even more
unstable modes. The positions of the mode frequencies $\lambda$ in the
complex plane vary continuously with $\kappa$, and these are just
values of $\kappa$ where a complex conjugate pair of frequencies
crosses the real axis. The results of Hirschmann and Eardley confirm
and subsume collapse simulation results by Liebling and Choptuik
\cite{LieblingChoptuik} for the scalar-Brans-Dicke system, and
collapse and perturbative results on the axion-dilaton system by
Hamad\'e, Horne and Stewart \cite{HamadeHorneStewart}. Where the CSS
solution fails to be the critical solution, a DSS solution takes
over. In particular, for $\kappa=0$, the free complex scalar field,
the critical solution is just the real scalar field DSS solution of
Choptuik.

Liebling \cite{Liebling} has found initial data sets that find the CSS
solution for values of $\kappa$ (for example $\kappa=0$) where the
true critical solution is DSS. The complex scalar field in these data
sets is of the form $\phi(r)=\exp i\omega r$ times a slowly varying
function of $r$, for arbitrary $r$, while its momentum $\Pi(r)$ is
either zero or $d\phi/dr$. Conversely, data sets that are purely real
find the DSS solution even for values of $\kappa$ where the true
critical solution is the CSS solution, for example for
$\kappa=1$. These two special families of initial data maximize and
minimize the $U(1)$ charge. Small deviations from these data find the
sub-dominant ``critical'' solution for some time, then veer off and
find the true critical solution. (Even later, of course, the critical
solution is also abandoned in turn for dispersion or black hole
formation.)


\subsection{Approximate self-similarity and universality classes}


As we have seen, the presence of a length scale in the field equations
can give rise to static (or oscillating) asymptotically flat critical
solutions and a mass gap at the black hole threshold. Depending on the
initial data and on how the scale appears in the field equations, this
scale can also become asymptotically irrelevant as a self-similar
solution reaches ever smaller spacetime scales. This behavior was
already noticed by Choptuik in the collapse of a massive scalar field,
or one with an arbitrary potential term generally \cite{Choptuik94}
and confirmed by Brady, Chambers and Gon\c calves
\cite{BradyChambersGoncalves}. It was also seen in the spherically
symmetric EYM system \cite{ChoptuikChmajBizon}. In order to capture
the notion of an asymptotically self-similar solution, one may set the
arbitrary scale $L$ in the definition (\ref{x_tau}) of $\tau$ to the
scale set by the field equations, here $1/m$.

Introducing suitable dimensionless first-order variables $Z$ (such as
$a$, $\alpha$, $\phi$, $r\phi_{,r}$ and $r\phi_{,t}$ for the
spherically symmetric scalar field), one can write the field equations
as a first order system
\begin{equation}
F\left(Z,Z_{,x},Z_{,\tau},e^{-\tau}\right)=0.
\end{equation}
Every appearance of $m$ gives rise to an appearance of $e^{-\tau}$. If
the field equations contain only positive integer powers of $m$, one
can make an ansatz for the critical solution of the form
\begin{equation}
\label{asymptotic_CSS}
Z_*(x,\tau) = \sum_{n=0}^\infty e^{-n\tau} Z_n(x).
\end{equation}
This is an expansion around a scale-invariant solution $Z_0$ (obtained
by setting $m\to 0$, in powers of (scale on which the solution
varies)/(scale set by the field equations).

After inserting the ansatz into the field equations, each $Z_n(x)$ is
calculated recursively from the preceding ones. For large enough
$\tau$ (on spacetime scales small enough, close enough to the
singularity), this expansion is expected to converge. A similar ansatz
can be made for the linear perturbations of $Z_*$, and solved again
recursively. Fortunately, one can calculate the leading order
background term $Z_0$ on its own, and obtain the exact echoing period
$\Delta$ in the process (in the case of DSS). Similarly, one can
calculate the leading order perturbation term on the basis of $Z_0$
alone, and obtain the exact value of the critical exponent $\gamma$ in
the process. This procedure was carried out by Gundlach
\cite{Gundlach_EYM} for the Einstein-Yang-Mills system, and by
Gundlach and Mart\'\i n-Garc\'\i a \cite{GundlachMartin} for massless
scalar electrodynamics. Both systems have a single scale $1/e$ (in
units $c=G=1$), where $e$ is the gauge coupling constant.

The leading order term $Z_0$ in the expansion of the self-similar
critical solution $Z_*$ obeys the equation
\begin{equation}
F\left(Z_0,Z_{0,x},Z_{0,\tau},0\right)=0.
\end{equation}
Clearly, this leading order term is independent of the overall scale
$L$. The critical exponent $\gamma$ depends only on $Z_0$, and is
therefore also independent of $L$. There is a region in the space of
initial data where in fine-tuning to the black hole threshold the
scale $L$ becomes irrelevant, and the behaviour is dominated by the
critical solution $Z_0$. In this region, the usual type II critical
phenomena occur, independently of the value of $L$ in the field
equations. In this sense, all systems with a single length scale $L$
in the field equations are in one universality class
\cite{HaraKoikeAdachi,GundlachMartin}.  The massive scalar field, for
any value of $m$, or massless scalar electrodynamics, for any value of
$e$, are in the same universality class as the massless scalar field.

It should be stressed that universality classes with respect to a
dimensionful parameter arise in regions of phase space (which may be
large). Another region of phase space may be dominated by an
intermediate attractor that has a scale proportional to $L$. This is
the case for the massive scalar field with mass $m$: in one region of
phase space, the black hole threshold is dominated by the Choptuik
solution and type II critical phenomena occur, in another, it is
dominated by metastable oscillating boson stars, whose mass is $1/m$
times a factor of order 1 \cite{BradyChambersGoncalves}.

This notion of universality classes is fundamentally the same as in
statistical mechanics. Other examples include modifications to the
perfect fluid equation of state that do not affect the limit of high
density. The $SU(2)$ Yang-Mills and $SU(2)$ Skyrme models, in
spherical symmetry, also belong to the same universality class
\cite{BizonChmajTabor}.

If there are several scales $L_0$, $L_1$, $L_2$ etc. present in the
problem, a possible approach is to set the arbitrary scale in
(\ref{x_tau}) equal to one of them, say $L_0$, and define the
dimensionless constants $l_i=L_i/L_0$ from the others.  The size of
the universality classes depends on where the $l_i$ appear in the
field equations. If a particular $L_i$ appears in the field equations
only in positive integer powers, the corresponding $l_i$ appears only
multiplied by $e^{-\tau}$, and will be irrelevant in the scaling
limit. All values of this $l_i$ therefore belong to the same
universality class. As an example, adding a quartic self-interaction
$\lambda\phi^4$ to the massive scalar field, for example, gives rise
to the dimensionless number $\lambda/m^2$, but its value is an
irrelevant (in the language of renormalization group theory)
parameter. All self-interacting scalar fields are in fact in the same
universality class. Contrary to the statement in
\cite{GundlachMartin}, I would now conjecture that massive scalar
electrodynamics, for any values of $e$ and $m$, forms a single
universality class in a region of phase space where type II critical
phenomena occur. Examples of dimensionless parameters which do change
the universality class are the $k$ of the perfect fluid, the $\kappa$
of the 2-dimensional sigma model, or a conformal coupling of the
scalar field.


\subsection{Gravity regularizes self-similar matter} 
\label{subsection:regular}


One important aspect of self-similar critical solutions is that they
have no equivalent in the limit of vanishing gravity. The critical
solution arises from a time evolution of smooth, even analytic initial
data. It should therefore itself be analytic outside the future of its
singularity. Self-similar spherical matter fields in spacetime are
singular either at the center of spherical symmetry (to the past of
the singularity), or at the past characteristic cone of the
singularity. Only adding gravity makes solutions possible that are
regular at both places. As an example we consider the spherical
massless scalar field.


\subsubsection{The massless scalar field on flat spacetime}


It is instructive to consider the self-similar solutions of a simple
matter field, the massless scalar field, in spherical symmetry without
gravity. The general solution of the spherically symmetric wave
equation is of course
\begin{equation} 
\phi(r,t) = r^{-1}\left[f(t+r)-g(t-r)\right],
\end{equation}
where $f(z)$ and $g(z)$ are two free functions of one variable ranging
from $-\infty$ to $\infty$. $f$ describes ingoing and $g$ outgoing
waves. Regularity at the center $r=0$ for all $t$ requires $f(z)=g(z)$
for $f(z)$ a smooth function. Physically this means that ingoing waves
move through the center and become outgoing waves. Now we transform to
new coordinates $x$ and $\tau$ defined by
\begin{equation}
\label{global_CSS}
r = e^{-\tau} \cos x, \qquad t = e^{-\tau} \sin x,
\end{equation}
and with range $-\infty<\tau<\infty$, $-\pi/2\le x\le \pi/2$. These
coordinates are adapted to self-similarity, but unlike the $x$ and
$\tau$ introduced in (\ref{x_tau}) they cover all of Minkowski space
with the exception of the point $(t=r=0)$. The general solution of the
wave equation for $t>r$ can formally be written as
\begin{eqnarray}
\phi(r,t)=\phi(x,\tau) & = & (\tan x + 1)F_+\left[\ln(\sin x + \cos x)
- \tau\right] \nonumber \\ & - & (\tan x - 1)G_+\left[\ln(\sin x -
\cos x) - \tau\right],
\end{eqnarray}
through the substitution $f(z)/z=F_+(\ln z)$ and $g(z)/z=G_+(\ln z)$
for $z>0$. Similarly, we define $f(z)/z=F_-[\ln (-z)]$ and
$g(z)/z=G_-[\ln (-z)]$ for $z<0$ to cover the sectors $|t|<r$ and
$t<-r$. Note that $F_+(z)$ and $F_-(z)$ together contain the same
information as $f(z)$.

Continuous self-similarity $\phi=\phi(x)$ is equivalent to $F_\pm(z)$
and $G_\pm(z)$ being constant. Discrete self-similarity requires them
to be periodic in $z$ with period $\Delta$. The condition for
regularity at $r=0$ for $t>0$ is $F_+=G_+$, while regularity at $r=0$
for $t<0$ requires $F_-=G_-$. Regularity at $t=r$ requires $G_\pm$ to
vanish, while regularity at $t=-r$ requires $F_\pm$ to vanish.

We conclude that a self-similar solution (continuous or discrete), is
either zero everywhere, or else it is regular in only one of three
places: at the center $r=0$ for $t\ne 0$, at the past light cone
$t=-r$, or at the future light cone $t=r$. We conjecture that other
simple matter fields, such as the perfect fluid, show similar
behavior.


\subsubsection{The self-gravitating massless scalar field}


The presence of gravity changes this singularity structure
qualitatively. Dimensional analysis applied to the metric
(\ref{CSS_coordinates}) or (\ref{DSS_coordinates}) shows that
$\tau=\infty$ [the point $(t=r=0)$] is now a curvature singularity
(unless the self-similar spacetime is Minkowski). But elsewhere, the
solution can be more regular. There is a one-parameter family of exact
spherically symmetric scalar field solutions found by Roberts
\cite{Roberts} that is regular at both the future and past light cone
of the singularity, not only at one of them. (It is singular at the
past and future branch of $r=0$.) The only solution without gravity
with this property is $\phi=0$. The Roberts solution will be discussed
in more detail in section \ref{subsection:singularity} below.

Similarly, the scale-invariant or scale-periodic solutions found in
near-critical collapse simulations are regular at both the past branch
of $r=0$ and the past light cone (or sound cone, in the case of the
perfect fluid). Once more, in the absence of gravity only the trivial
solution has this property.

I have already argued that the critical solution must be as smooth on
the past light cone as elsewhere, as it arises from the collapse of
generic smooth initial data. No lowering of differentiability or other
unusual behavior should take place before a curvature singularity
arises at the center. As Evans first realized, this requirement turns
the scale-invariant or scale-periodic ansatz into a boundary value
problem between the past branch of $r=0$ and the past sound cone, that
is, roughly speaking, between $x=0$ and $x=1$.

In the CSS ansatz in spherical symmetry suitable for the perfect
fluid, all fields depend only on $x$, and one obtains an ODE boundary
value problem. In a scale-periodic ansatz in spherical symmetry, such
as for the scalar field, all fields are periodic in $\tau$, and one
obtains a 1+1 dimensional hyperbolic boundary value problem on a
coordinate square, with regularity conditions at, say, $x=0$ and
$x=1$, and periodic boundary conditions at $\tau=0$ and
$\tau=\Delta$. Well-behaved numerical solutions of these problems have
been obtained, with numerical evidence that they are locally unique,
and they agree well with the universal solution that emerges in
collapse simulations (references are given in the column ``Critical
solution'' of Table \ref{table:mattermodels}). It remains an open
mathematical problem to prove existence and (local) uniqueness of the
solution defined by regularity at the center and the past light cone.

One important technical detail should be mentioned here. In the curved
solutions, the past light cone of the singularity is not in general
$r=-t$, or $x=1$, but is given by $x=x_0$, or in the case of
scale-periodicity, by $x=x_0(\tau)$, with $x_0$ periodic in $\tau$ and
initially unknown. The same problem arises for the sound cone. It is
convenient to make the coordinate transformation
\begin{equation}
\bar x = {x \over x_0(\tau)}, \qquad \bar \tau = {2\pi\over
\Delta}\tau,
\end{equation}
so that the sound cone or light cone is by definition at $\bar x =1$,
while the origin is at $\bar x =0$, and so that the period in
$\bar\tau$ is now always $2\pi$. In the DSS case the periodic function
$x_0(\bar\tau)$ and the constant $\Delta$ now appear explicitly in the
field equations, and they must be solved for as nonlinear eigenvalues.
In the CSS case, the constant $x_0$ appears, and must be solved for as
a nonlinear eigenvalue.

As an example for a DSS ansatz, we give the equations for the
spherically symmetric massless scalar field in the coordinates
(\ref{x_tau}) adapted to self-similarity and in a form ready for
posing the boundary value problem. (The equations of
\cite{Gundlach_Chop1} have been adapted to the notation of this
review.) We introduce the first-order matter variables
\begin{equation}
\label{Xpm}
X_\pm = \sqrt{2\pi} r\left({\phi_{,r}\over a}\pm {\phi_{,t}\over
\alpha}\right),
\end{equation}
which describe ingoing and outgoing waves. It is also useful to
replace $\alpha$ by
\begin{equation}
D = \left(1-{\Delta\over2\pi}{d\ln x_0\over d\bar\tau}\right) {xa\over
\alpha}
\end{equation}
as a dependent variable.  In the scalar field wave equation
(\ref{wave}) we use the Einstein equations (\ref{dalpha_dr}) and
(\ref{da_dt}) to eliminate $a_{,t}$ and $\alpha_{,r}$, and obtain
\begin{equation}
\bar x {\partial X_\pm \over \partial \bar x} = (1 \mp D)^{-1} \left\{
\left[{1\over2}(1-a^2)-a^2X_\mp^2\right] X_\pm - X_\mp \pm D
\left({\Delta\over2\pi}-{d\ln x_0\over d\bar\tau}\right)^{-1}
{\partial X_\pm\over\partial \bar\tau}\right\}.
\end{equation}
The three Einstein equations (\ref{da_dr},\ref{dalpha_dr},\ref{da_dt})
become
\begin{eqnarray}
{\bar x\over a}{\partial a\over\partial\bar x} & = & {1\over2}(1-a^2)
+{1\over2} a^2 (X_+^2+X_-^2), \\ {\bar x\over D}{\partial
D\over\partial\bar x} & = & 2 - a^2, \\ 0 & = & (1-a^2) +a^2
(X_+^2+X_-^2) - a^2 D^{-1}(X_+^2-X_-^2) \nonumber \\ && +
\left({\Delta\over2\pi}-{d\ln x_0\over d\bar\tau}\right)^{-1} {2\over
a} {\partial a\over \partial \bar\tau} .
\end{eqnarray}
As suggested by the format of the equations, they can be treated as
four evolution equations in $\bar x$ and one constraint that is
propagated by them. The freedom in $x_0(\bar\tau)$ is to be used to
make $D=1$ at $\bar x=1$. Now $\bar x=0$ and $\bar x=1$ resemble
``regular singular points'', if we are prepared to generalize this
concept from linear ODEs to nonlinear PDEs. Near $\bar x=0$, the four
evolution equations are clearly of the form $\partial Z/\partial \bar
x = {\rm regular}/\bar x$. That $\bar x=1$ is also a regular singular
point becomes clearest if we replace $D$ by $\bar D=(1-D)/(\bar x-1)$.
The ``evolution'' equation for $X_+$ near $\bar x=1$ then takes the
form $\partial X_+/\partial \bar x = {\rm regular}/(\bar x -1)$, while
the other three equations are regular.

This format of the equations also demonstrates how to restrict from a
DSS to a CSS ansatz: one simply drops the $\bar \tau$-derivatives. The
constraint then becomes algebraic, and the resulting ODE system can be
considered to have three rather than four dependent variables.

Given that the critical solutions are regular at the past branch of
$r=0$ and at the past sound cone of the singularity, and that they are
self-similar, one would expect them to be singular at the future light
cone of the singularity (because after solving the boundary value
problem there is no free parameter left in the solution). The real
situation is more subtle as we shall see in Section
\ref{subsection:singularity}.


\subsection{Critical phenomena and naked singularities}
\label{subsection:singularity}


Choptuik's results have an obvious bearing on the issue of cosmic
censorship. (For a general review of cosmic censorship, see
\cite{Wald_censorship}.) As we shall see in this section, the critical
spacetime has a naked singularity. This spacetime can be approximated
arbitrarily well up to fine-tuning of a generic parameter. A region of
arbitrarily high curvature is seen from infinity as fine-tuning is
improved. Critical collapse therefore provides a set of smooth initial
data for naked singularity formation that has codimension one in phase
space. It does not violate cosmic censorship if one states it as
``generic(!) smooth initial data for reasonable matter do not form
naked singularities''.

Nevertheless, critical collapse is an interesting test of cosmic
censorship. First of all, the set of data is of codimension one,
certainly in the space of spherical asymptotically flat data, and
apparently \cite{Gundlach_nonspherical} also in the space of all
asymptotically flat data. This means that one can fine-tune any
generic parameter, whichever comes to hand, as long as it
parameterizes a smooth curve in the space of initial data. Secondly,
critical phenomena seem to be generic with respect to matter models,
including realistic matter models with intrinsic scales. These two
features together mean that, in a hypothetical experiment to create a
Planck-sized black hole in the laboratory through a strong explosion,
one could fine-tune any one design parameter of the bomb, without
requiring control over its detailed effects on the explosion.

The metric of the critical spacetime is of the form $e^{-2\tau}$ times
a regular metric. From this general form alone, one can conclude that
$\tau=\infty$ is a curvature singularity, where Riemann and Ricci
invariants blow up like $e^{4\tau}$, and which is at finite proper
time from regular points.  The Weyl tensor with index position
${C^a}_{bcd}$ is conformally invariant, so that components with this
index position remain finite as $\tau\to\infty$. In this property it
resembles the initial singularity in Penrose's Weyl tensor conjecture
rather than the final singularity in generic gravitational
collapse. This type of singularity is called ``conformally
compactifiable'' \cite{Tod_pc} or ``isotropic'' \cite{Goodeetal}. Is
the singularity naked, and is it timelike, null or a ``point''? The
answer to these questions remains confused, partly because of
coordinate complications, partly because of the difficulty of
investigating the singular behavior of solutions numerically.

Choptuik's, and Evans and Coleman's, numerical codes were limited to
the region $t<0$, in the Schwarzschild-like coordinates
(\ref{tr_metric}), with the origin of $t$ adjusted so that the
singularity is at $t=0$. Evans and Coleman conjectured that the
singularity is shrouded in an infinite redshift based on the fact that
$\alpha$ grows as a small power of $r$ at constant $t$. This is
directly related to the fact that $a$ goes to a constant $a_\infty>1$
as $r\to\infty$ at constant $t$, as one can see from the Einstein
equation (\ref{dalpha_dr}). This in turn means simply that the
critical spacetime is not asymptotically flat, but asymptotically
conical at spacelike infinity, with the Hawking mass proportional to
$r$.

Hamad\'e and Stewart \cite{HamadeStewart} evolved near-critical scalar
field spacetimes on a double null grid, which allowed them to follow
the time evolution up to close to the future light cone of the
singularity. They found evidence that this light cone is not preceded
by an apparent horizon, that it is not itself a (null) curvature
singularity, and that there is only a finite redshift along outgoing
null geodesics slightly preceding it. (All spherically symmetric
critical spacetimes appear to be qualitatively alike as far as the
singularity structure is concerned, so that what we say about one is
likely to hold for the others.)

Hirschmann and Eardley \cite{HE1} were the first to continue a
critical solution itself right up to the future light cone. They
examined a CSS complex scalar field solution that they had constructed
as a nonlinear ODE boundary value problem, as discussed in Section
\ref{subsection:regular}. (This particular one is not a proper
critical solution, but that should not matter for the global
structure.) They continued the ODE evolution in the self-similar
coordinate $x$ through the coordinate singularity at $t=0$ up to the
future light cone by introducing a new self-similarity coordinate $x$.
The self-similar ansatz reduces the field equations to an ODE
system. The past and future light cones are regular singular points of
the system, at $x=x_1$ and $x=x_2$. At these ``points'' one of the two
independent solutions is regular and one singular. The boundary value
problem that originally defines the critical solution corresponds to
completely suppressing the singular solution at $x=x_1$ (the past
light cone). The solution can be continued through this point up to
$x=x_2$. There it is a mixture of the regular and the singular
solution.

We now state this more mathematically. The ansatz of Hirschmann and
Eardley for the self-similar complex scalar field is (we slightly
adapt their notation)
\begin{equation}
\phi(x,\tau) = f(x) e^{i\omega\tau}, \quad a=a(x), \quad
\alpha=\alpha(x),
\end{equation}
with $\omega$ a real constant. Near the future light cone they find
that $f$ is approximately of the form
\begin{equation}
f(x)\simeq C_{\rm reg}(x) + (x-x_2)^{(i\omega+1)(1+\epsilon)} C_{\rm
sing}(x),
\end{equation}
with $C_{\rm reg}(x)$ and $C_{\rm sing(x)}$ regular at $x=x_2$, and
$\epsilon$ a small positive constant. The singular part of the scalar
field oscillates an infinite number of times as $x\to x_2$, but with
decaying amplitude. This means that the scalar field $\phi$ is just
differentiable, and that therefore the stress tensor is just
continuous. It is crucial that spacetime is not flat, or else
$\epsilon$ would vanish. For this in turn it is crucial that the
regular part $C_{\rm reg}$ of the solution does not vanish, as one
sees from the field equations.

The only other case in which the critical solution has been continued
up to the future light cone is Choptuik's real scalar field solution
\cite{Gundlach_Chop2}. Let $X_+$ and $X_-$ be the ingoing and outgoing
wave degrees of freedom respectively defined in (\ref{Xpm}). At the
future light cone $x=x_2$ the solution has the form
\begin{eqnarray}
X_-(x,\tau) & \simeq & f_-(x,\tau), \\ X_+(x,\tau) & \simeq &
f_+(x,\tau) + (x-x_2)^\epsilon f_{\rm sing}(x,\tau - C\ln x) ,
\end{eqnarray}
where $C$ is a positive real constant, $f_-$, $f_+$ and $f_{\rm sing}$
are regular real functions with period $\Delta$ in their second
argument, and $\epsilon$ is a small positive real constant. (We have
again simplified the original notation.) Again, the singular part of
the solution oscillates an infinite number of times but with decaying
amplitude. Gundlach concludes that the scalar field, the metric
coefficients, all their first derivatives, and the Riemann tensor
exist, but that is as far as differentiability goes. (Not all second
derivatives of the metric exist, but enough to construct the Riemann
tensor.) If either of the regular parts $f_-$ or $f_+$ vanished,
spacetime would be flat, $\epsilon$ would vanish, and the scalar field
itself would be singular. In this sense, gravity regularizes the
self-similar matter field ansatz. In the critical solution, it does
this perfectly at the past lightcone, but only partly at the future
lightcone. Perhaps significantly, spacetime is almost flat at the
future horizon in both the examples, in the sense that the Hawking
mass divided by $r$ is a very small number. In the spacetime of
Hirschmann and Eardley it appears to be as small as $10^{-6}$, but not
zero according to numerical work by Horne \cite{Horne_pc}.

In summary, the future light cone (or Cauchy horizon) of these two
critical spacetimes is not a curvature singularity, but it is singular
in the sense that differentiability is lower than elsewhere in the
solution. Locally, one can continue the solution through the future
light cone to an almost flat spacetime (the solution is of course not
unique). It is not clear, however, if such a continuation can have a
regular center $r=0$ (for $t>0$), although this seems to have been
assumed in \cite{HE1}. A priori, one should expect a conical
singularity, with a (small) defect angle at $r=0$.

The results just discussed were hampered by the fact that they are
investigations of singular spacetimes that are only known in numerical
form, with a limited precision. As an exact toy model we consider an
exact spherically symmetric, CSS solution for massless real scalar
field that was apparently first discovered by Roberts \cite{Roberts}
and then re-discovered in the context of critical collapse by Brady
\cite{Brady_Roberts} and Oshiro et al. \cite{Oshiro_Roberts}. We use
the notation of Oshiro et al. The solution can be given in double null
coordinates as
\begin{eqnarray}
\label{Roberts1}
ds^2 & = & - du\,dv + r^2(u,v) \,d\Omega^2, \\ r^2(u,v) & = & {1\over
4}\left[(1-p^2)v^2 - 2vu + u^2\right], \\
\label{Roberts3}
\phi(u,v) & = & {1\over 2} \ln {(1-p)v - u\over (1+p) v - u},
\end{eqnarray}
with $p$ a constant parameter. (Units $G=c=1$.) Two important
curvature indicators, the Ricci scalar and the Hawking mass, are
\begin{equation}
R={p^2 uv\over 2r^4}, \quad M = - {p^2 uv\over 8r}.
\end{equation}
The center $r=0$ has two branches, $u=(1+p)v$ in the past of $u=v=0$,
and $u=(1-p)v$ in the future.  For $0<p<1$ these are timelike
curvature singularities. The singularities have negative mass, and the
Hawking mass is negative in the past and future light cones. One can
cut these regions out and replace them by Minkowski space, not
smoothly of course, but without creating a $\delta$-function in the
stress-energy tensor. The resulting spacetime resembles the critical
spacetimes arising in gravitational collapse in some respects: it is
self-similar, has a regular center $r=0$ at the past of the curvature
singularity $u=v=0$ and is continuous at the past light cone. It is
also continuous at the future light cone, and the future branch of
$r=0$ is again regular.

It is interesting to compare this with the genuine critical solutions
that arise as attractors in critical collapse. They are as regular as
the Roberts solution (analytic) at the past $r=0$, more regular
(analytic versus continuous) at the past light cone, as regular
(continuous) at the future light cone and, it is to be feared, less
regular at the future branch of $r=0$: In contrary to previous claims
\cite{HE1,Gundlach_Banach} there may be no continuation through the
future sound or light cone that does not have a conical singularity at
the future branch of $r=0$. The global structure still needs to be
clarified for all known critical solutions.

In summary, the critical spacetimes that arise asymptotically in the
fine-tuning of gravitational collapse to the black-hole threshold have
a curvature singularity that is visible at infinity with a finite
redshift. The Cauchy horizon of the singularity is mildly singular
(low differentiability), but the curvature is finite there. It is
unclear at present if the singularity is timelike or if there exists a
continuation beyond the Cauchy horizon with a regular center, so that
the singularity is limited, loosely speaking, to a point. Further work
should be able to clarify this. In any case, the singularity is naked
and the critical solutions therefore provide counter-examples to any
formulation of cosmic censorship which states only that naked
singularities cannot arise from smooth initial data in reasonable
matter models. The statement must be that there is no {\it open ball}
of smooth initial for naked singularities.

Recent analytic work by Christodoulou on the spherical scalar field
\cite{Christodoulou5} is not directly relevant to the smooth (analytic
or $C^\infty$) initial data discussed here. Christodoulou considers a
larger space of initial data that are not $C^1$. He shows that for any
data set $f_0$ in this class that forms a naked singularity there are
data $f_1$ and $f_2$ such that the data sets $f_0 + c_1 f_1 + c_2 f_2$
do not contain a naked singularity, for any $c_1$ and $c_2$ except
zero. Here $f_1$ is data of bounded variation, and $f_2$ is absolutely
continuous data. Therefore, the set of naked singularity data is at
least codimension two in the space of data of bounded variation, and
of codimension at least one in the space of absolutely continuous
data. The semi-numerical result of Gundlach claims that it is
codimension exactly one in the set of smooth data. The result of
Christodoulou holds for any $f_0$, including initial data for the
Choptuik solution. The apparent contradiction is resolved if one notes
that the $f_1$ and $f_2$ of Christodoulou are not smooth in (at least)
one point, namely where the initial data surface is intersected by the
past light cone of the singularity in $f_0$. (Roughly speaking, $f_1$
and $f_2$ start throwing scalar field matter into the naked
singularity at the exact moment it is born, and therefore depend on
$f_0$.) The data $f_0 + c_1 f_1 + c_2 f_2$ are therefore not smooth.


\subsection{Beyond spherical symmetry} \label{subsection:nonspherical}


Every aspect of the basic scenario: CSS and DSS, universality and
scaling applies directly to a critical solution that is not
spherically symmetric, but all the models we have described are
spherically symmetric. There are only two exceptions to date: a
numerical investigation of critical collapse in axisymmetric pure
gravity \cite{AbrahamsEvans}, and studies of the nonspherical
perturbations the spherically symmetric perfect fluid
\cite{Gundlach_nonspherical} and scalar field \cite{MartinGundlach}
critical solutions.  They correspond to two related questions: Are the
critical phenomena in the known spherically symmetric examples
destroyed already by small deviations from spherical symmetry? And:
are there critical phenomena in gravitational collapse far from
spherical symmetry?


\subsubsection{Axisymmetric gravitational waves}
\label{subsection:axisymm}


The paper of Abrahams and Evans \cite{AbrahamsEvans} was the first
paper on critical collapse to be published after Choptuik's PRL, but
it remains the only one to investigate a non-spherically symmetric
situation, and therefore also the only one to investigate critical
phenomena in the collapse of gravitational waves in vacuum. Because of
its importance, we summarize its contents here with some technical
detail.

The physical situation under consideration is axisymmetric vacuum
gravity. The numerical scheme uses a 3+1 split of the spacetime. The
ansatz for the spacetime metric is
\begin{equation}
\label{axisymmetric}
ds^2=-\alpha^2\,dt^2 + \phi^4\left[e^{2\eta/3}(dr+\beta^r\,dt)^2 +
r^2e^{2\eta/3}(d\theta+\beta^\theta\,dt)^2 +
e^{-4\eta/3}r^2\sin^2\theta\,d\varphi^2\right],
\end{equation}
parameterized by the lapse $\alpha$, shift components $\beta^r$ and
$\beta^\theta$, and two independent coefficients $\phi$ and $\eta$ in
the 3-metric. All are functions of $r$, $t$ and $\theta$. The fact
that $dr^2$ and $r^2\,d\theta^2$ are multiplied by the same
coefficient is called quasi-isotropic spatial gauge. The variables for
a first-order-in-time version of the Einstein equations are completed
by the three independent components of the extrinsic curvature,
$K^r_\theta$, $K^r_r$, and $K^\varphi_\varphi$.  The ansatz limits
gravitational waves to one ``polarisation'' out of two, so that there
are as many physical degrees of freedom as in a single wave
equation. In order to obtain initial data obeying the constraints,
$\eta$ and $K^r_\theta$ are given as free data, while the remaining
components of the initial data, namely $\phi$, $K^r_r$, and
$K^\varphi_\varphi$, are determined by solving the Hamiltonian
constraint and the two independent components of the momentum
constraint respectively.  There are five initial data variables, and
three gauge variables. Four of the five initial data variables, namely
$\eta$, $K^r_\theta$, $K^r_r$, and $K^\varphi_\varphi$, are updated
from one time step to the next via evolution equations. As many
variables as possible, namely $\phi$ and the three gauge variables
$\alpha$, $\beta^r$ and $\beta^\theta$, are obtained at each new time
step by solving elliptic equations. These elliptic equations are the
Hamiltonian constraint for $\phi$, the gauge condition of maximal
slicing (${K_i}^i=0$) for $\alpha$, and the gauge conditions
$g_{\theta\theta}=r^2 g_{rr}$ and $g_{r\theta}=0$ for $\beta^r$ and
$\beta^\theta$ (quasi-isotropic gauge).

For definiteness, the two free functions, $\eta$ and $K^r_\theta$, in
the initial data were chosen to have the same functional form they
would have in a linearized gravitational wave with pure $(l=2,m=0)$
angular dependence. Of course, depending on the overall amplitude of
$\eta$ and $K^r_\theta$, the other functions in the initial data will
deviate more or less from their linearized values, as the non-linear
initial value problem is solved exactly. In axisymmetry, only one of
the two degrees of freedom of gravitational waves exists. In order to
keep their numerical grid as small as possible, Abrahams and Evans
chose the pseudo-linear waves to be purely ingoing. (In nonlinear
general relativity, no exact notion of ingoing and outgoing waves
exists, but this ansatz means that the wave is initially ingoing in
the low-amplitude limit.) This ansatz (pseudo-linear, ingoing, $l=2$),
reduced the freedom in the initial data to one free function of
advanced time, $I^{(2)}(v)$. (In the linear limit, everything would
then remain a function of advanced time forever.) A suitably peaked
function was chosen.

Limited numerical resolution (numerical grids are now two-dimensional,
not one-dimensional as in spherical symmetry) allowed Abrahams and
Evans to find black holes with masses only down to $0.2$ of the ADM
mass. Even this far from criticality, they found power-law scaling of
the black hole mass, with a critical exponent $\gamma\simeq
0.36$. Determining the black hole mass is not trivial, and was done
from the apparent horizon surface area, and the frequencies of the
lowest quasi-normal modes of the black hole.  There was tentative
evidence for scale echoing in the time evolution, with $\Delta\simeq
0.6$, with about three echos seen. This corresponds to a scale range
of about one order of magnitude. By a lucky coincidence, $\Delta$ is
much smaller than in all other examples, so that several echos could
be seen without adaptive mesh refinement. The paper states that the
function $\eta$ has the echoing property $\eta(e^\Delta r,e^\Delta
t)=\eta(r,t)$. If the spacetime is DSS in the sense defined above, the
same echoing property is expected to hold also for $\alpha$, $\phi$,
$\beta^r$ and $r^{-1}\beta^\theta$, as one sees by applying the
coordinate transformation (\ref{x_tau}) to (\ref{axisymmetric}).

In a subsequent paper \cite{AbrahamsEvans2}, universality of the
critical solution, echoing period and critical exponent was
demonstrated through the evolution of a second family of initial data,
one in which $\eta=0$ at the initial time. In this family, black hole
masses down to $0.06$ of the ADM mass were achieved.  Further work on
critical collapse far away from spherical symmetry would be desirable,
but appears to be held up by numerical difficulty.


\subsubsection{Perturbing around spherical symmetry}


A different, and technically simpler, approach is to take a known
critical solution in spherical symmetry, and perturb it using
nonspherical perturbations.  Addressing this perturbative question,
Gundlach \cite{Gundlach_nonspherical} has studied the generic
non-spherical perturbations around the critical solution found by
Evans and Coleman \cite{EvansColeman} for the $p={1\over3}\rho$
perfect fluid in spherical symmetry. There is exactly one spherical
perturbation mode that grows towards the singularity (confirming the
previous results \cite{KoikeHaraAdachi,Maison}). There are no growing
nonspherical modes at all. A corresponding result was established for
non-spherical perturbations of the Choptuik solution for the massless
scalar field \cite{MartinGundlach}.

The main significance of this result, even though it is only
perturbative, is to establish one critical solution that really has
only one unstable perturbation mode within the full phase space.  As
the critical solution itself has a naked singularity (see Section
\ref{subsection:singularity}), this means that there is, for this
matter model, a set of initial data of codimension one in the full
phase space of general relativity that forms a naked singularity.
This result also confirms the role of critical collapse as the most
``natural''way of creating a naked singularity.


\subsection{Black hole charge and angular momentum}


Given the scaling power law for the black hole mass in critical
collapse, one would like to know what happens if one takes a generic
one-parameter family of initial data with both electric charge and
angular momentum (for suitable matter), and fine-tunes the parameter
$p$ to the black hole threshold. Does the mass still show power-law
scaling? What happens to the dimensionless ratios $L/M^2$ and $Q/M$,
with $L$ the black hole angular momentum and $Q$ its electric charge?
Tentative answers to both questions have been given using
perturbations around spherically symmetric uncharged collapse.


\subsubsection{Charge}


Gundlach and Mart\'\i n-Garc\'\i a \cite{GundlachMartin} have studied
scalar massless electrodynamics in spherical symmetry. Clearly, the
real scalar field critical solution of Choptuik is a solution of this
system too. Less obviously, it remains a critical solution within
massless (and in fact, massive) scalar electrodynamics in the sense
that it still has only one growing perturbation mode within the
enlarged solution space. Some of its perturbations carry electric
charge, but as they are all decaying, electric charge is a subdominant
effect. The charge of the black hole in the critical limit is
dominated by the most slowly decaying of the charged modes. From this
analysis, a universal power-law scaling of the black hole charge
\begin{equation}
Q\sim (p-p_*)^\delta
\end{equation}
was predicted. The predicted value $\delta\simeq 0.88$ of the critical
exponent (in scalar electrodynamics) was subsequently verified in
collapse simulations by Hod and Piran \cite{HodPiran_charge}. (The
mass scales with $\gamma\simeq 0.37$ as for the uncharged scalar
field.) General considerations using dimensional analysis led Gundlach
and Mart\'\i n-Garc\'\i a to the general prediction that the two
critical exponents are always related, for any matter model, by the
inequality
\begin{equation}
\delta\ge2\gamma.
\end{equation}
This has not yet been verified in any other matter model.


\subsubsection{Angular momentum}


Gundlach's results on non-spherically symmetric perturbations around
spherical critical collapse of a perfect fluid
\cite{Gundlach_nonspherical} allow for initial data, and therefore
black holes, with infinitesimal angular momentum.  All nonspherical
perturbations decrease towards the singularity. The situation is
therefore similar to scalar electrodynamics versus the real scalar
field. The critical solution of the more special model (here, the
strictly spherically symmetric fluid) is still a critical solution
within the more general model (a slightly nonspherical and slowly
rotating fluid). In particular, axial perturbations (also called
odd-parity perturbations) with angular dependence $l=1$ (i.e. dipole)
will determine the angular momentum of the black hole produced in
slightly supercritical collapse. Using a perturbation analysis similar
to that of Gundlach and Mart\'\i n-Garc\'\i a \cite{GundlachMartin},
Gundlach \cite{Gundlach_angmom} (see correction in
\cite{Gundlach_critfluid2}) has derived the angular momentum scaling
law
\begin{equation}
\label{CSS_L}
L \sim (p-p_*)^\mu
\end{equation}
For the range $0.123< k < 0.446$ of equations of state, the angular
momentum exponent $\mu$ is related to the mass exponent $\gamma$ by
\begin{equation}
\label{muofk}
\mu(k) = {5(1+3k)\over 3(1+k)} \gamma(k) .
\end{equation}
In particular for the value $k=1/3$, $\mu=(5/2)\gamma\simeq 0.898$.
An angular momentum exponent $\mu\simeq 0.76$ was derived for the
massless scalar field in \cite{GarfinkleGundlachMartin} using
second-order perturbation theory. Both results have not yet been
tested against numerical collapse simulations.


\section{Aspects of current research}
\label{section:otheraspects}


Like the previous section, this one contains extensions of the basic
ideas, but here we group together topics that are still under active
investigation. 


\subsection{Phase diagrams}
\label{subsection:phasediagrams}


In analogy with critical phenomena in statistical mechanics, let us
call a graph of the black hole threshold in the phase space of some
self-gravitating system a phase diagram. The full phase space is
infinite-dimensional, but one can plot a two-dimensional
submanifold. In such a plot the black hole threshold is generically a
line, analogous to the fluid/gas dividing line in the
pressure/temperature plane.

Interesting phenomena can be expected in systems that admit more
complicated phase diagrams. The massive complex scalar field for
example, admits stable stars as well as black holes and flat space as
possible end states. There are three phase boundaries, and these
should intersect somewhere. A generic two-parameter family of initial
data is expected to intersect each boundary in a line, and the three
lines should meet at a triple point.

Similarly, many systems admit both type I and type II phase
transitions, for example the massive real scalar field, and the
$SU(2)$ Yang-Mills field in spherical symmetry. In a two-dimensional
family of initial data, these should again generically show up as
lines, and generically these lines should intersect. Is the black hole
mass at the intersection finite or zero? Is there a third line that
begins where the type I and type II lines meet?

Choptuik, Hirschmann and Marsa \cite{ChoptuikHirschmannMarsa} have
investigated this for a specific two-parameter family of initial data
for the spherical $SU(2)$ Yang-Mills field, using a numerical
evolution code that can follow the time evolutions for long after a
black hole has formed. As known previously, the type I phase
transition is mediated by the static Bartnik-McKinnon solution, which
has one growing perturbation mode. The type II transition is mediated
by a DSS solution with one growing mode.  There is a third type of
phase transition along a third line which meets the intersection of
the type I and type II lines. On both sides of this ``type III'' phase
transition the final state is a Schwarzschild black hole with zero
Yang-Mills field strength, but the final state is distinguished by the
value of the Yang-Mills gauge potential at infinity. (The system has
two distinct vacuum states.)  The critical solution is an unstable
black hole with Yang-Mills hair, which collapses to a hairless
Schwarzschild black hole with either vacuum state of the Yang-Mills
field, depending on the sign of its one growing perturbation mode. The
critical solution is not unique, but is a member of a 1-parameter
family of hairy black holes parameterized by their mass. At the triple
point the family ends in a zero mass black hole.


\subsection{The renormalisation group as a time
evolution} \label{subsection:coordinates}


It has been pointed out by Argyres \cite{Argyres}, Koike, Hara and
Adachi \cite{KoikeHaraAdachi} and others that the time evolution near
the critical solution can be considered as a renormalisation group
flow on the space of initial data. The calculation of the critical
exponent in section \ref{subsection:scaling} is in fact mathematically
identical to that of the critical exponent governing the correlation
length near the critical point in statistical mechanics
\cite{Yeomans}, if one identifies the time evolution in the time
coordinate $\tau$ and spatial coordinate $x$ with the renormalisation
group flow. But those coordinates were defined only on self-similar
spacetimes plus linear perturbations. In order to obtain a full
renormalisation group, one has to generalize them to arbitrary
spacetimes, or, in other words, find a general prescription for the
lapse and shift as functions of arbitrary Cauchy data.

For simple parabolic or hyperbolic differential equations, a discrete
renormalisation (semi)group acting on their solutions has been defined
in the following way
\cite{Goldenfeld,BricmontKupiainen,ChenGoldenfeld,ChenGoldenfeldOOno}. Evolve
initial data over a certain finite time interval, then rescale the
final data in a certain way. Solutions which are fixed points under
this transformation are scale-invariant, and may be attractors. One
nice distinctive feature of GR as opposed to these simple models is
that one can use a shift freedom in GR (one that points inward towards
an accumulation point) to incorporate the rescaling into the time
evolution, and the lapse freedom to make each rescaling by a constant
factor an evolution through a constant time ($\tau$, in our notation)
interval.

The crucial distinctive feature of general relativity, however, is
that a solution does not correspond to a unique trajectory in the
space of initial data. This is because a spacetime can be sliced in
different ways, and on each slice one can have different coordinate
systems. Infinitesimally, this slicing and coordinate freedom is
parameterized by the lapse and shift.  In a relaxed notation, one can
write the ADM equations as $(\dot g,\dot K)=\rm{functional}
(g,K,\alpha,\beta)$, where $g$ is the 3-metric, $K$ the extrinsic
curvature, $\alpha$ the lapse and $\beta$ the shift.  The lapse and
shift can be set freely, independently of the initial data. Of course
they influence only the coordinates on the spacetime, not the
spacetime itself, but the ADM equations are not yet a dynamical
system. If we specify a prescription $(\alpha,\beta)=\rm{functional}
(g,K)$, then substituting it into the ADM equations, we obtain $(\dot
g,\dot K)=\rm{functional}(g,K)$, which is an (infinite-dimensional)
dynamical system.  We are then faced with the general question: Given
initial data in general relativity, is there a prescription for the
lapse and shift, such that, if these are in fact data for a
self-similar solution, the resulting time evolution actively drives
the metric to the special form (\ref{DSS_coordinates}) that explicitly
displays the self-similarity?

An algebraic prescription for the lapse suggested by Garfinkle
\cite{Garfinkle2} did not work, but maximal slicing with zero shift
does work \cite{GarfinkleMeyer} if combined with a manual rescaling of
space. Garfinkle and Gundlach \cite{GarfinkleGundlach} have suggested
several combinations of lapse and shift conditions that not only leave
CSS spacetimes invariant, but also turn the Choptuik DSS spacetime
into a limit cycle. The combination of maximal slicing with minimal
strain shift has the nice property that it also turns static
spacetimes into fixed points (and probably periodic spacetimes into
limit cycles). Maximal slicing requires the first slice to be maximal
(${K_a}^a=0$), but other prescriptions allow for an arbitrary initial
slice with arbitrary spatial coordinates. All these coordinate
conditions are elliptic equations that require boundary conditions,
and will turn CSS spacetimes into fixed points only for correct
boundary conditions. Roughly speaking, these boundary conditions
require a guess of how far the slice is from the accumulation point
$t=t_*$, and answers to this problem only exist in spherical symmetry.


\subsection{Analytic approaches} 


A number of authors have attempted to explain critical collapse with
the help of analytic solutions. The one-parameter family of exact
self-similar real massless scalar field solutions first discovered by
Roberts \cite{Roberts} has already been presented in section
\ref{subsection:singularity}. It has been discussed in the context of
critical collapse in \cite{Brady_Roberts,Oshiro_Roberts}, and later
\cite{WangOliveira,Burko}. The original, analytic, Roberts solution is
cut and pasted to obtain a new solution which has a regular center
$r=0$ and which is asymptotically flat. Solutions from this family
[see Eqns. (\ref{Roberts1}-\ref{Roberts3})] with $p>1$ can be
considered as black holes, and to leading order around the critical
value $p=1$, their mass is $M\sim(p-p_*)^{1/2}$. The pitfall in this
approach is that only perturbations within the self-similar family are
considered, so the formal critical exponent applies only to this one,
very special, family of initial data. But the $p=1$ solution has many
growing perturbations which are spherically symmetric (but not
self-similar), and is therefore not a critical solution in the sense
of being an attractor of codimension one. This was already clear
because it did not appear in collapse simulations at the black hole
threshold, but Frolov has calculated the perturbation spectrum
analytically \cite{Frolov,Frolov3}. The eigenvalues of spherically
symmetric perturbations fill a sector of the complex plane, with
$Re\lambda\le1$ . All nonspherical perturbations decay.  Other
supposed critical exponents that have been derived analytically are
usually valid only for a single, very special family of initial data
also.

Other authors have employed analytic approximations to the actual
Choptuik solution.  Pullin \cite{Pullin_Chop} has suggested describing
critical collapse approximately as a perturbation of the Schwarzschild
spacetime. Price and Pullin \cite{PricePullin} have approximated the
Choptuik solution by two flat space solutions of the scalar wave
equation that are matched at a ``transition edge'' at constant
self-similarity coordinate $x$. The nonlinearity of the gravitational
field comes in through the matching procedure, and its details are
claimed to provide an estimate of the echoing period $\Delta$. While
the insights of this paper are qualitative, some of its ideas reappear
in the construction \cite{Gundlach_Chop1} of the Choptuik solution as
a 1+1 dimensional boundary value problem. Frolov \cite{Frolov4} has
suggested approximating the Choptuik solution as the Roberts solution
plus its most rapidly growing (spherical) perturbation mode, pointing
out that it oscillates in $\tau$ with a period $4.44$, but ignoring
the fact that also grows exponentially. This is probably a misguided
approach.

In summary, purely analytic approaches have so far remained
unsuccessful in explaining critical collapse.


\subsection{Astrophysical black holes}


Any real world application of critical phenomena would require that
critical phenomena are not an artifact of the simple matter models
that have been studied so far, and that they are not an artifact of
spherical symmetry. At present this seems a reasonable hypothesis.

Critical collapse still requires a kind of fine-tuning of initial data
that does not seem to arise naturally in the astrophysical
world. Niemeyer and Jedamzik \cite{NiemeyerJedamzik} have suggested a
scenario that gives rise to such fine-tuning. In the early universe,
quantum fluctuations of the metric and matter can be important, for
example providing the seeds of galaxy formation. If they are large
enough, these fluctuations may even collapse immediately, giving rise
to what is called primordial black holes. Large quantum fluctuations
are exponentially more unlikely than small ones, $P(\delta)\sim
\exp-\delta^2$, where $\delta$ is the density contrast of the
fluctuation. One would therefore expect the spectrum of primordial
black holes to be sharply peaked at the minimal $\delta$ that leads to
black hole formation. That is the required fine-tuning. In the
presence of fine-tuning, the black hole mass is much smaller than the
initial mass of the collapsing object, here the density
fluctuation. In consequence, the peak of the primordial black hole
spectrum might be expected to be at exponentially smaller values of
the black hole mass than expected naively. See also
\cite{NiemeyerJedamzik2,Yokoyama}.

The primordial black holes work assumes that the critical phenomena
will be of type II. If one could fine-tune the gravitational collapse
of stars made of realistic matter (i.e. not scalar fields) it seems
likely that type I critical phenomena could be observed, i.e. there
would be a universal mass gap. Critical collapse is not likely to be
relevant in the real universe (at least at present) as there is no
mechanism for fine-tuning of initial data.


\subsection{Critical collapse in semiclassical gravity}


As we have seen in the last section, critical phenomena may provide a
natural route from everyday scale down to much smaller scales, perhaps
down to the Planck scale. Various authors have investigated the
relationship of Choptuik's critical phenomena to quantum black
holes. It is widely believed that black holes should emit thermal
quantum radiation, from considerations of quantum field theory on a
fixed Schwarzschild background on the one hand, and from the purely
classical three laws of black hole mechanics on the other (see
\cite{Wald_BH} for a review). But there is no complete model of the
back-reaction of the radiation on the black hole, which should be
shrinking. In particular, it is unknown what happens at the endpoint
of evaporation, when full quantum gravity should become important. It
is debated in particular if the information that has fallen into the
black hole is eventually recovered in the evaporation process or lost.

To study these issues, various 2-dimensional toy models of gravity
coupled to scalar field matter have been suggested which are more or
less directly linked to a spherically symmetric 4-dimensional
situation (see \cite{Giddings_BH} for a review). In two space-time
dimensions, the quantum expectation value of the matter stress tensor
can be determined from the trace anomaly alone, together with the
reasonable requirement that the quantum stress tensor is
conserved. Furthermore, quantizing the matter scalar field(s) $f$ but
leaving the metric classical can be formally justified in the limit of
many such matter fields. The two-dimensional gravity used is not the
two-dimensional version of Einstein gravity but of a scalar-tensor
theory of gravity. $e^\phi$, where $\phi$ is called the dilaton, in
the 2-dimensional toy model plays essentially the role of $r$ in 4
spacetime dimensions. There seems to be no preferred 2-dimensional toy
model, with arbitrariness both in the quantum stress tensor and in the
choice of the classical part of the model. In order to obtain a
resemblance of spherical symmetry, a reflecting boundary condition is
imposed at a timelike curve in the 2-dimensional spacetime. This plays
the role of the curve $r=0$ in a 2-dimensional reduction of the
spherically symmetric 4-dimensional theory.

How does one expect a model of semiclassical gravity to behave when
the initial data are fine-tuned to the black hole threshold?  First of
all, until the fine-tuning is taken so far that curvatures on the
Planck scale are reached during the time evolution, universality and
scaling should persist, simply because the theory must approximate
classical general relativity. Approaching the Planck scale from above,
one would expect to be able to write down a critical solution that is
the classical critical solution asymptotically at large scales, as an
expansion in inverse powers of the Planck length.  This ansatz would
recursively solve a semiclassical field equation, where powers of
$e^{\tau}$ (in coordinates $x$ and $\tau$) signal the appearances of
quantum terms.  Note that this is exactly the ansatz
(\ref{asymptotic_CSS}), but with the opposite sign in the exponent, so
that the higher order terms now become negligible as $\tau\to-\infty$,
that is away from the singularity on large scales. On the Planck scale
itself, this ansatz would not converge, and self-similarity would
break down.

Addressing the question from the side of classical general relativity,
Chiba and Siino \cite{ChibaSiino} write down a 2-dimensional toy
model, and add a quantum stress tensor that is determined by the trace
anomaly and stress-energy conservation. They note that the quantum
stress tensor diverges at $r=0$. Ayal and Piran \cite{AyalPiran} make
an ad-hoc modification to these semiclassical equations. They modify
the quantum stress tensor by a function which interpolates between 1
at large $r$, and $r^2/L_p^2$ at small $r$. They justify this
modification by pointing out that the resulting violation of energy
conservation takes place only at the Planck scale. It takes place,
however, not only where the solution varies dynamically on the Planck
scale, but at all times in a Planck-sized world tube around the center
$r=0$, even before the solution itself reaches the Planck scale
dynamically. This introduces a non-geometric, background structure,
effect at the world-line $r=0$. With this modification, Ayal and Piran
obtain results in agreement with our expectations set out above. For
far supercritical initial data, black formation and subsequent
evaporation are observed. With fine-tuning, as long as the solution
stays away from the Planck scale, critical solution phenomena
including the Choptuik universal solution and critical exponent are
observed. (The exponent is measured as $0.409$, indicating a limited
accuracy of the numerical method.) In an intermediate regime, the
quantum effects increase the critical value of the parameters
$p$. This is interpreted as the initial data partly evaporating while
they are trying to form a black hole.

Researchers coming from the quantum field theory side seem to favor a
model (the RST model) in which ad hoc ``counter terms'' have been
added to make it soluble. The matter is a conformally rather than
minimally coupled scalar field. The field equations are trivial up to
an ODE for a timelike curve on which reflecting boundary conditions
are imposed. The world line of this ``moving mirror'' is not clearly
related to $r$ in a 4-dimensional spherically symmetric model, but
seems to correspond to a finite $r$ rather than $r=0$. This may
explain why the problem of a diverging quantum stress tensor is not
encountered. Strominger and Thorlacius \cite{StromingerThorlacius}
find a critical exponent of $1/2$, but their 2-dimensional situation
differs from the 4-dimensional one in many aspects. Classically
(without quantum terms) any ingoing matter pulse, however weak, forms
a black hole. With the quantum terms, matter must be thrown in
sufficiently rapidly to counteract evaporation in order to form a
black hole. The initial data to be fine-tuned are replaced by the
infalling energy flux. There is a threshold value of the energy flux
for black hole formation, which is known in closed form. (Recall this
is a soluble system.) The mass of the black hole is defined as the
total energy it absorbs during its lifetime.  This black hole mass is
given by
\begin{equation}
M\simeq \left({\delta\over\alpha}\right)^{1\over2}
\end{equation}
where $\delta$ is the difference between the peak value of the flux
and the threshold value, and $\alpha$ is the quadratic order
coefficient in a Taylor expansion in advanced time of the flux around
its peak. There is universality with respect to different shapes of
the infalling flux in the sense that only the zeroth and second order
Taylor coefficients matter.

Peleg, Bose and Parker \cite{PelegBoseParker,BoseParkerPeleg} study
the so-called CGHS 2-dimensional model. This (non-soluble) model does
allow for a study of critical phenomena with quantum effects turned
off. Again, numerical work is limited to integrating an ODE for the
mirror world line. Numerically, the authors find black hole mass
scaling with a critical exponent of $\gamma\simeq 0.53$. They find the
critical solution and the critical solution to be universal with
respect to families of initial data. Turning on quantum effects, the
scaling persists to a point, but the curve of $\ln M$ versus
$\ln(p-p_*)$ then turns smoothly over to a horizontal
line. Surprisingly, the value of the mass gap is not universal but
depends on the family of initial data. While this is the most
``satisfactory'' result among those discussed here from the classical
point of view, one should keep in mind that all these results are
based on mere toy models of quantum gravity.

Rather than using a consistent model of semiclassical gravity, Brady
and Ottewill \cite{BradyOttewill} calculate the quantum stress-energy
tensor of a conformally coupled scalar field on the fixed background
of the perfect fluid CSS critical solution and treat it as an
additional perturbation, on top of the perturbations of the fluid-GR
system itself. In doing this, they neglect the coupling between fluid
and quantum scalar perturbations through the metric
perturbations. From dimensional analysis, the quantum perturbation has
a Lyapunov exponent $\lambda=2$. If this is larger than the positive
Lyapunov exponent $\lambda_0$, it will become the dominant
perturbation for sufficiently good fine-tuning, and therefore
sufficiently good fine-tuning will reveal a mass gap. For a spherical
perfect fluid with equation of state $p=k\rho$, one finds that
$\lambda_0>2$ for $k>0.53$, and vice versa. If $\lambda_0>2$, the
semiclassical approximation breaks down for sufficiently good
fine-tuning, and this calculation remains inconclusive.


\section{Conclusions}
\label{section:conclusions}


We conclude with separate summings-up of what is known today and what
still needs to be investigated and understood.


\subsection{Summary}


When one fine-tunes a smooth one-parameter family of smooth,
asymptotically flat initial data to get close enough to the black hole
threshold, the details of the initial data are completely forgotten in
a small spacetime region where the curvature is high, and all
near-critical time evolutions converge to one universal solution
there. (This region is limited both in space and time, and at late
times the final state is either a black hole or empty space.) At the
black hole threshold, there either is a universal minimum black hole
mass (type I transition), or black hole formation starts at
infinitesimal mass (type II transition). In a type I transition, the
universal critical solution is time-independent, or periodic in time,
and the closer the initial data are to the black hole threshold, the
longer it persists. In a type II transition, the universal critical
solution is scale-invariant or scale-periodic, and the closer the
initial data are to the black hole threshold, the smaller the black
hole mass, by the famous formula (\ref{power_law}).

Both types of behavior arise because there is a solution which is an
intermediate attractor, or attractor of codimension one. Its basin of
attraction is the black hole threshold itself, a hypersurface of
codimension one that bisects phase space.  Any time evolution that
begins with initial data near the black hole threshold (but not
necessarily close to the critical solution) first approaches the
critical solution, then moves away from it along its one growing
perturbation mode. At late times, the solution only remembers on which
side of the black hole threshold the initial data were, and how far
away from the threshold.

Our understanding of critical phenomena rests on this dynamical
systems picture, but crucial details of the picture have not yet been
defined rigorously. Nevertheless, it suggests semi-analytic
perturbative calculations that have been successful in predicting the
scaling of black hole mass and charge in critical collapse to high
precision.

The importance of type II behavior lies in providing a natural route
from large (the initial data) to arbitrarily small (the final black
hole) scales, with possible applications to astrophysics and quantum
gravity. Fine-tuning any one generic parameter in the initial data to
the black hole threshold, for a number of matter models, without
assuming any other symmetries, will do the trick.

Type II critical behavior also clarifies what version of cosmic
censorship one can hope to prove. At least in some matter models
(scalar field, perfect fluid), fine-tuning any smooth one-parameter
family of smooth, asymptotically flat initial data, without any
symmetries, gives rise to a naked singularity. In this sense the set
of initial data that form a naked singularity is codimension one in
the full phase space of smooth asymptotically flat initial data for
well-behaved matter. Any statement of cosmic censorship in the future
can only exclude naked singularities arising from {\it generic}
initial data.

Finally, critical phenomena are arguably the outstanding contribution
of numerical relativity to knowledge in GR to date, and they continue
to act as a motivation and a source of testbeds for numerical
relativity.


\subsection{Outlook}


Clearly, more numerical work will be useful to further establish the
generality of critical phenomena in gravitational collapse, or to find
a counter-example instead. In particular, future research should
include highly non-spherical situations, initial data with large
angular momentum and/or electric charge, and matter models with a
large number of internal degrees of freedom (for example,
collisionless matter instead of a perfect fluid). Both going beyond
spherical symmetry and including collisionless matter pose formidable
numerical challenges.

The fundamental theoretical challenge is to explain why so many matter
models admit a critical solution, that is, an attractor of codimension
one at the black hole threshold. If the existence of a critical
solution is really a generic feature, then there should be at least an
intuitive argument, and perhaps a mathematical proof, for this
important fact.  On the other hand, the spherical Einstein-Vlasov
system may already be providing a counter-example. A more thorough
mathematical and numerical investigation of this system is therefore
particularly urgent.

The critical spacetimes and their perturbations are well known only in
the past light cone of the singularity.  The Cauchy horizon and the
naked singularity itself, as well as the possible continuations beyond
the Cauchy horizon, of the critical spacetimes have not yet been
investigated thoroughly. It is unknown if all possible continuations
have a timelike naked singularity, and in what manner this singularity
is avoided when one perturbs away from the black hole threshold.

An important mathematical challenge is to make the intuitive dynamical
systems picture of critical collapse more rigorous, by providing a
distance measure on the phase space, and a prescription for a flow on
the phase space (equivalent to a prescription for the lapse and
shift). The latter problem is intimately related to the problem of
finding good coordinate systems for the binary black hole problem.

On the phenomenological side, it is likely that the scope of critical
collapse will be expanded to take into account new phenomena, such as
multicritical solutions (with several growing perturbation modes), or
critical solutions that are neither static, periodic, CSS or DSS. More
complicated phase diagrams than the simple black
hole-dispersion transition are already being examined, and the
intersections of phase boundaries are of particular interest.


\subsection{Thanks}


A large number of people have contributed indirectly to this paper,
but I would particularly like to thank Pat Brady, Matt Choptuik, David
Garfinkle, Jos\'e M. Mart\'\i n-Garc\'\i a, Alan Rendall and (last but
not least) Bob Wald for stimulating discussions on many aspects of
critical collapse.



\begin{figure}
\epsfysize=6cm
\centerline{\epsffile{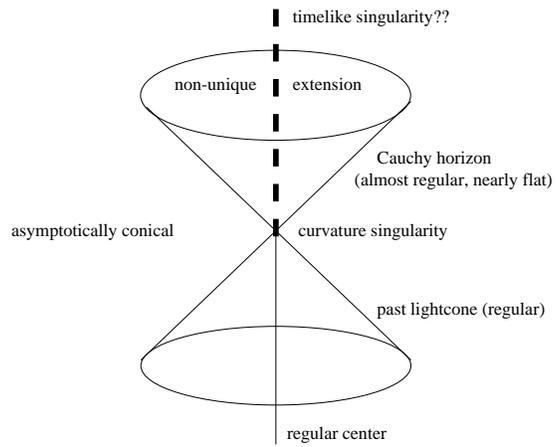}}
\caption{The global structure of spherically symmetric critical spacetimes. One
dimension in spherical symmetry has been suppressed.} 
\end{figure}


\end{document}